\documentclass[lettersize,journal]{IEEEtran}
\usepackage{algorithm}
\usepackage{array}
\usepackage[caption=false,font=normalsize,labelfont=sf,textfont=sf]{subfig}
\usepackage{textcomp}
\usepackage{stfloats}
\usepackage{url}
\usepackage{verbatim}
\usepackage{graphicx}
\usepackage{algpseudocode} 
\usepackage{cite}
\usepackage{amsmath,amssymb,amsfonts}
\usepackage{booktabs,caption}
\usepackage[flushleft]{threeparttable}

\usepackage{placeins}

\usepackage{textcomp}
\usepackage{psfrag}
\usepackage{subfig}
\usepackage{caption}
\usepackage{romannum}

\usepackage{multirow}
\usepackage{textcomp}
\usepackage[table,xcdraw]{xcolor}
\usepackage{tabto}
\usepackage{colortbl}
\usepackage{hhline}
\usepackage{cite}
\usepackage{url}
\usepackage{verbatim}
\usepackage[export]{adjustbox}
\usepackage{pdfpages}
\begin{document}
\title{Neuromorphic Retina: An FPGA-based Emulator}

\author{Prince Phillip$^{\ast}$,~\IEEEmembership{Student Member,~IEEE,}
        Pallab Kumar Nath$^{\ast}$,~\IEEEmembership{Member,~IEEE,}
        Kapil Jainwal,~\IEEEmembership{Member,~IEEE,}\\
       André van Schaik,~\IEEEmembership{Fellow,~IEEE}
	and~Chetan Singh Thakur,~\IEEEmembership{Senior Member,~IEEE}

\thanks{Prince Philip and Chetan Singh Thakur
are with the NeuRonICS Lab, Department of Electronic Systems Engineering, Indian Institute of Science, Bangalore, KA, 560012 INDIA (e-mail: princephilip@iisc.ac.in, csthakur@iisc.ac.in).}%
\thanks{Pallab Kumar Nath was with the NeuRonICS Lab, Department of Electronic Systems Engineering, Indian Institute of Science and is currently associated with the Department of Electronics Communication Engineering, Pandit Deendayal Energy University, Gandhinagar email: (pallab.nath@sot.pdpu.ac.in)}
\thanks{Kapil Jainwal is with the Department of  Electrical Engineering, Indian Institute of Technology, Hyderabad, 502285, INDIA (email: kapiljainwal@ee.iith.ac.in).}
\thanks{André van Schaik is with the International Centre for Neuromorphic Systems, The MARCS Institute for Brain, Behaviour, and Development, Western Sydney University, Sydney, NSW 2751, AUSTRALIA (email: a.vanschaik@westernsydney.edu.au).}
\thanks{$^{\ast}$ These authors have made equal contributions to this paper
}
}

\maketitle
\begin{abstract}
Implementing accurate models of the retina is a challenging task, particularly in the context of creating visual prosthetics and devices. Notwithstanding the presence of diverse artificial renditions of the retina, the imperative task persists to pursue a more realistic model. In this work, we are emulating a neuromorphic retina model on an FPGA. The key feature of this model is its powerful adaptation to luminance and contrast, which allows it to accurately emulate the sensitivity of the biological retina to changes in light levels. Phasic and tonic cells are realizable in the retina in the simplest way possible. Our FPGA implementation of the proposed biologically inspired digital retina, incorporating a receptive field with a center-surround structure, is reconfigurable and can support 128$\times$128 pixel images at a frame rate of 200fps. It consumes 1720 slices, approximately 3.7k Look-Up Tables (LUTs), and Flip-Flops (FFs) on the FPGA. This implementation provides a high-performance, low-power, and small-area solution and could be a significant step forward in the development of biologically plausible retinal prostheses with enhanced information processing capabilities. 
\end{abstract}


\section{Introduction}
The retina assumes a fundamental role in the visual pathway, acting as the initial site where visual stimuli are captured, encoded, and initiating the intricate process of visual perception. The photoreceptor cells in the retina constantly convert light into electrical signals. These signals then trigger a series of chemical reactions within retinal neurons, resulting in the generation of electrical current in the form of nerve impulses or action potentials. These electrical impulses travel along the optic nerve to the lateral geniculate nucleus (LGN) in the thalamus. From the LGN, the visual information is relayed to the primary visual cortex, also known as V1 or the striate cortex located in the occipital lobe of the brain. Subsequently, the visual signals propagate to advanced visual processing areas in the brain for further analysis and interpretation. It has been discovered in neurobiological investigations that the retina executes complicated processing of visual information, such as brightness computation, motion, and edge recognition \cite{masland2011cell,masland2012neuronal,field2007information,gollisch2010eye,levick1967receptive,roska2014retina,sanes2015types}. Thus, accurate mapping of the retina's input to output is vital for clear vision.
 
 \begin{figure}[t]
	\centering
	\subfloat[]{\includegraphics[width=0.45\linewidth]{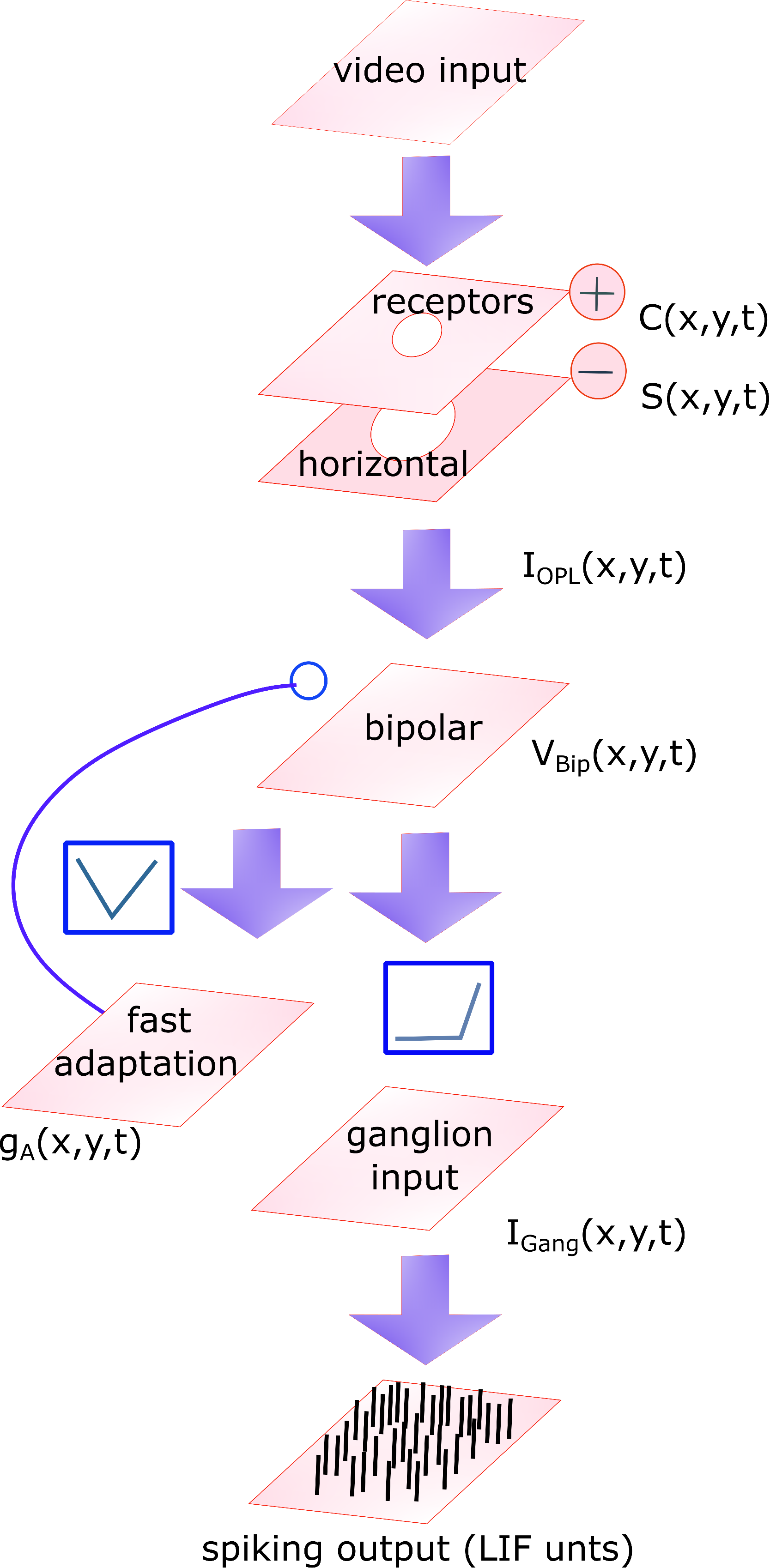}
		\label{retina_model}}
  \hspace{0.2cm}
	\subfloat[]{\includegraphics[width=0.2\linewidth]{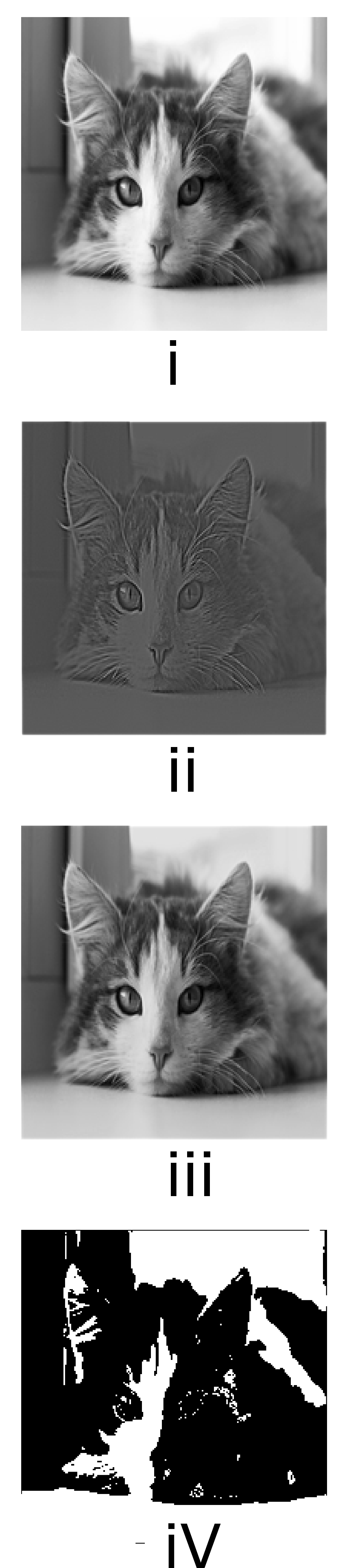}
		\label{software_op}}
\caption{\protect\subref{retina_model}~Simplified schema showing the retina implementation in the digital realm, starting from photoreceptor cell receiving external video stimuli from the outside environment to the spike response generated by the ganglion cells;~\protect\subref{software_op}~Images depicting the effect of stimulus at different retina layer model where, (\romannum{1}) shows the $10^{th}$ frame of an input video stimulus, (\romannum{2}) shows the impact of the OPL layer output, (\romannum{3}) illustrates the result of the bipolar layer, and (\romannum{4}) shows the resultant spiking activity.}
	\label{architecture}
\end{figure}

Several studies have reported on the hardware implementation of different features of the biological retina. A mammalian retina model with its emulated digital architecture is reported in \cite{nagy2005emulated}. They utilized a digital cellular neural network to examine the behavior of the outer layer of the retina. Specialized processors for emulating different retina models are described in \cite{voroshazi2008advanced}, and \cite{voroshazi2009fpga}. This architecture is useful for rapid model building and parameter tuning in retina modeling. In  \cite{yang2019digital}, and \cite{wang2011numerical}, the spiking pattern of a retinal network under light stimulation has been investigated and implemented on an FPGA. They demonstrate that hardware is capable of producing retinal ganglion cell responses. Also, some research has been done on the signal-flow platform of the retina \cite{eshraghian2016modelling,eshraghian2017biological,ghanbarpour2021efficient}. However, these digital models do not capture the key retinal features such as the receptive field (RF) center-surround structure observed in retinal ganglion cells (RGCs), spatio-temporal filtering, adaptation to luminance, contrast gain control, phasic cells (transient), and tonic cells (sustained) generate an output that is close to the biological retina.  

The center-surround receptive field structure is essential for lateral inhibition to occur \cite{warren2016mechanisms,meister1999neural}. Lateral inhibition is important because it is the ability of the sensory systems to improve the perception of the edges of stimuli. Luminance and contrast-gain control are important retinal adaptations present in all species \cite{whitmire2016rapid,ghodrati2019contrast}. These mechanisms enable the retina to accommodate a broad input range from the visual system, spanning over ten orders of magnitude of illumination. This occurs within the confined output range of optic nerve fibers, which can only generate up to hundreds of impulses per second. Throughout the visual system, luminance adaptation and contrast-gain control modify the temporal dynamics and neural response gain to preserve perceptual sensitivity under various illumination conditions \cite{ohzawa1985contrast,shapley1984visual,chander2001adaptation}. These adaptations enhance information transmission \cite{brenner2000adaptive,fairhall2001lewen} or feature detection and processing \cite{ringach2007operating,sharpee2006adaptive}. These mechanisms also have a significant impact on brightness \cite{wangnew}.  Tonic cells are important for recognizing the presence of visual input and preserving a consistent representation of it throughout time \cite{wohrer2009virtual}. For activities like tracking moving objects, this can be helpful.
 The detection of changes in a visual signal, such as its beginning, offset, or movement, depends on phasic cells \cite{wohrer2009virtual}.
 Phasic cells can be helpful for activities like locating an object in a cluttered visual scene.

 This work introduces a novel bio-plausible digital retina, which is implemented based on the Convis retina model \cite{huth2018convis}. The key contributions of this work in relation to previous approaches are as follows:
\begin{itemize}

\item In the digital retina, the outer plexiform layer (OPL) filter exhibits non-separability in both spatial and temporal domains. Consequently, in a spatially uniform scene, this non-separable spatio-temporal filter can effectively detect changes in brightness, a capability not achievable with a separable filter.
    
    \item In biological retinas, the signal from the surround component of the OPL layer is sent with an additional delay. This latency is incorporated in the digital retina. The OPL layer is implemented as a difference of Gaussian (DoG) in terms of space and as a biphasic filter relating to time. Hence, the center-surround OPL filter of the digital retina performs the dual functions of an edge and a movement detector.

\item A high-pass filter at the photoreceptor level incorporates luminance adaptation.

\item The distinctive nonlinear dependency of the retinal filtering on the average contrast level is the model for the contrast gain control mechanism, which also accounts for the contrast gain control simultaneously connected to the temporal and spatial organization of the input. Signal shaping is carried out
 at the inner plexiform layer (IPL).

\item Phasic and tonic cells can be attained in the digital retina through the variation of the strength of a partially high-pass filter.
    \item The proposed digital retina is reconfigurable and processes 128$\times$128 pixel frames at a frame rate of 200fps.
    
\end{itemize}
The rest of the sections are organized as follows. Section \ref{archdigital} describes the digital retina architecture. Section \ref{fpgaimplementation} explains the FPGA implementation details of the retina. Section \ref{resultsdigital} presents the simulation results of the digital retina with brief discussions and a comparison with other related work. Finally, the paper is concluded in Section \ref{conclusiondigital}.
 \begin{figure*}[!t]%
    \centering
    \subfloat[\centering  ]{{\includegraphics[scale=0.17]{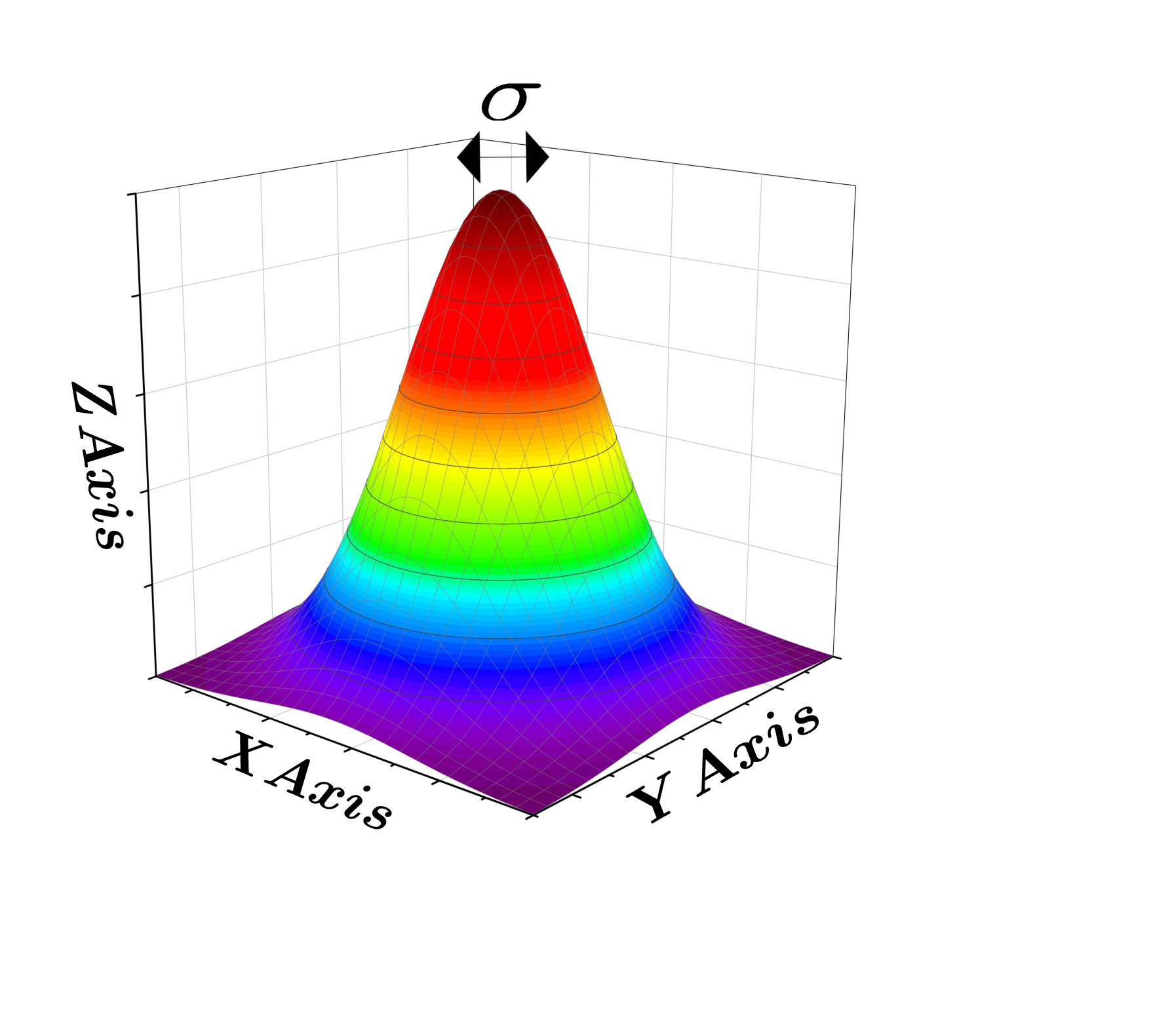} }%
    \label{spatialfilter}}
    \subfloat[\centering  ]{{\includegraphics[scale=0.23]{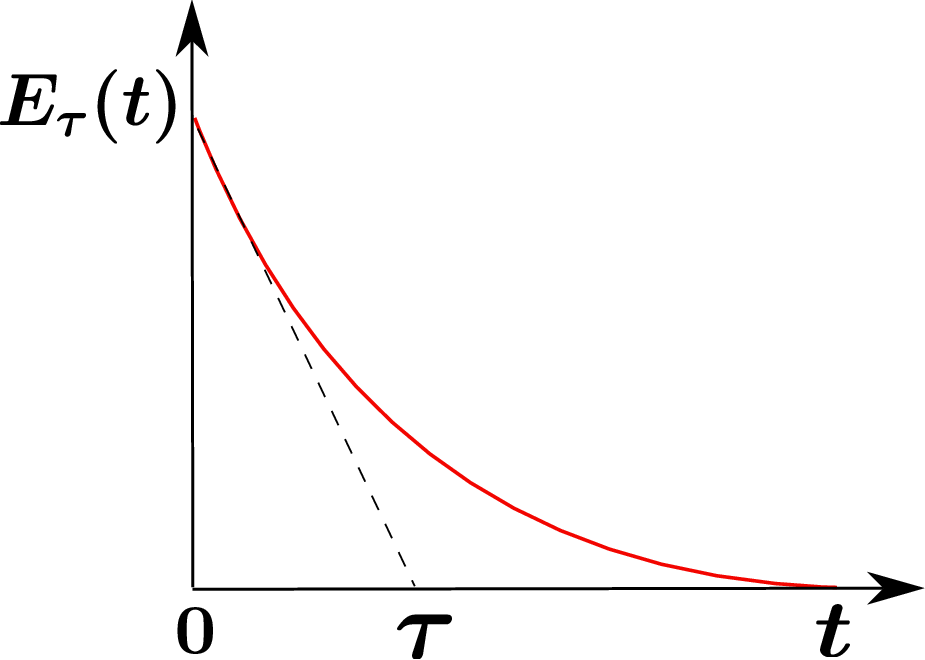} }%
    \label{lowpass}}
    \qquad
    \subfloat[\centering]
    {{\includegraphics[scale=0.23]{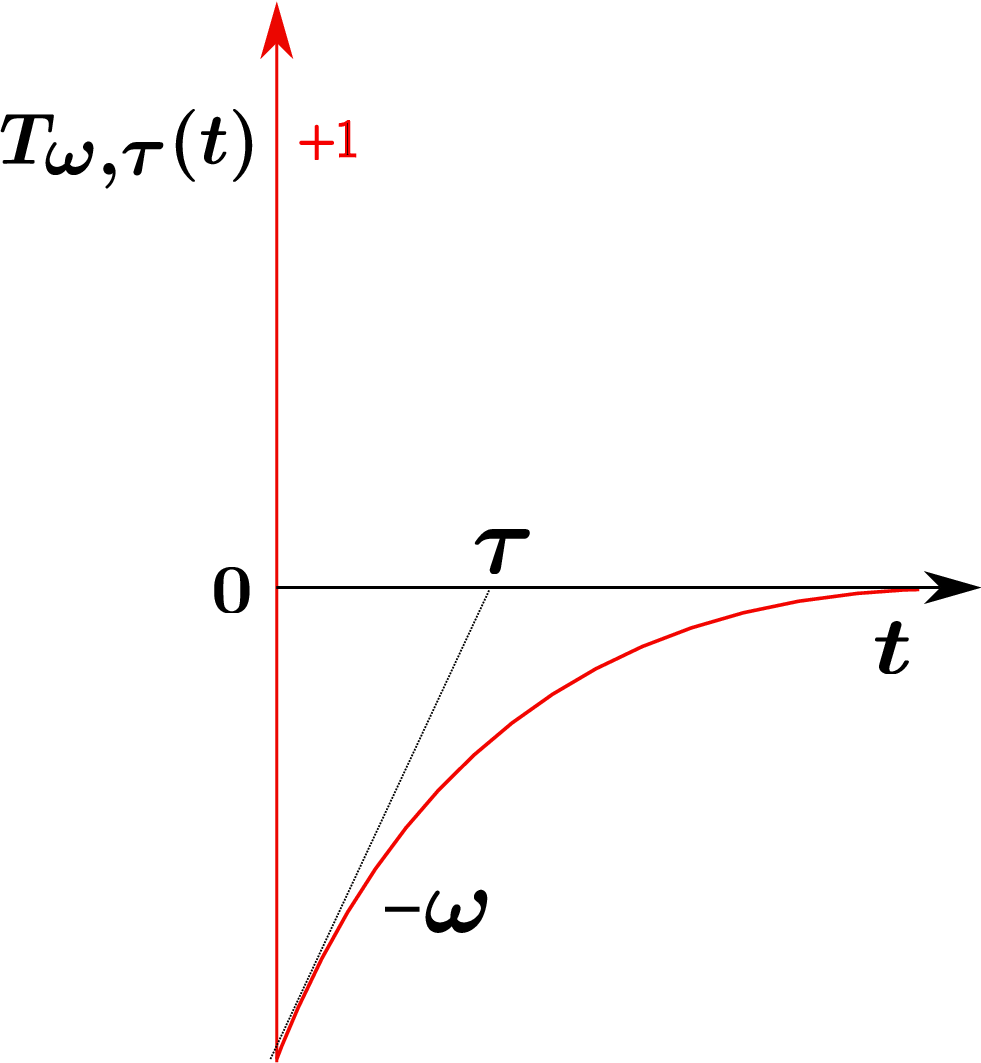} }%
    \label{highpass}}
    \qquad
    \caption{Kernels used in the neuromorphic digital retina for spatial and temporal filtering in the retinal layers;  \protect\subref{spatialfilter} Depicts spatial low-pass filter kernel (Gaussian) for $G_\text{C}$, $G_\text{S}$, and $G_\text{A}$; \protect\subref{lowpass} Shows temporal low-pass filter kernel for $E_{\tau \text{C}}$,  $E_{\tau \text{S}}$, and  $E_{\tau \text{A}}$; \protect\subref{highpass} Shows temporal high-pass filter kernel for $T_{\text{w},\tau}$ and $T_{G}$. Symbol $\sigma$ indicates the spatial constant, while $\tau$ represents the time constant.}%
    \label{kernels}%
\end{figure*}
\section{Digital Retina Architecture}
\label{archdigital}
The digital retina models neurons from photoreceptors to ganglion cells, as shown in Fig. \ref{retina_model}. In the outer plexiform layer (OPL), photoreceptors capture light, and horizontal cells create center-surround receptive fields through lateral inhibition. The output signal of the OPL layer undergoes gain control using bipolar-level nonlinear contrast gain control to adapt to different light contrast levels. Signal shaping occurs in the inner plexiform layer (IPL), where ganglion cell activity is modeled with leaky integrate-and-fire (LIF) neurons. Each stage is detailed in the following subsections
\subsection{Outer Plexiform Layer (OPL)}
In this retinal layer, synapses form connections between horizontal and photoreceptor cells, shaping a receptive field with a center-surround structure. The triad synapse within the OPL contributes to the center-surround antagonism observed in the bipolar cell layer. The generation of the center signal involves temporal low-pass, temporal high-pass, and spatial filters. The surround signal is obtained by delaying the center signal and applying additional spatial smoothing. Subtracting the surround signal ($S$) from the center signal ($C$) yields the OPL current ($I_\text{OPL}$). This center-surround interaction is crucial for the OPL layer's functionality and has been described in \cite{huth2018convis,wohrer2009virtual} as given by,
\begin{equation}
 C=G_\text{C}*T_{\text{w},\tau}* E_{\tau \text{C}}*L
 \label{ctr}
\end{equation}
\begin{equation}
  S=G_\text{S}*E_{\tau \text{S}}*C 
   \label{srd}
\end{equation}
 \begin{equation}
 I_\text{OPL}=\lambda_\text{OPL}(C - \omega_\text{OPL}S)   
\end{equation}
  where $L$ is the luminance input, $E_{\tau \text{C}}$ and $T_{\text{w},\tau}$ are temporal low-pass and high-pass filters at the center signal and $E_{\tau \text{S}}$ is a temporal low-pass filter for the surround signal. $G_\text{C}$ and $G_\text{S}$ are the spatial filters (Gaussian) for the center and surround signal.
  $\lambda_\text{OPL}$ and $\omega_\text{OPL}$ are scalar weights. The * indicates the convolution operation.
\subsubsection{Spatial filters}
In the retina, neurons are connected via gap junctions. In the outer plexiform layer (OPL), photoreceptors and horizontal cells are linked through these junctions, represented by spatial filters ($G_\text{C}$ and $G_\text{S}$) in Eqs. (\ref{ctr}) and (\ref{srd}). Horizontal cells, with broader receptive fields, use a higher spatial constant ($\sigma$), while photoreceptors, with smaller receptive fields, use a lower $\sigma$ to capture fine details. The spatial filter kernel is shown in Fig. \ref{spatialfilter}.
\subsubsection{Temporal filters}
Temporal filters in retina modeling capture and mimic neuron responses. They approximate temporal processing and enhance understanding of the retina's capabilities. Our implementation involves low-pass and high-pass filters.
\paragraph{Temporal low-pass filter}
Temporal low-pass filtering starts at photoreceptors, where light is converted to electrical signals through biochemical reactions. These signals experience synaptic delays and membrane integration in later retinal layers. Synaptic delay arises from neurotransmitter release, diffusion, and receptor activation. Integration involves summing and processing incoming signals, influenced by synaptic properties and cell membrane characteristics. The impulse response of a low-pass filter with time constant $\tau$ is described by the equation as
\begin{equation}
			      	E_{\tau}=\frac{exp(-(t/\tau)}{\tau}
			         \label{exp}    
         \end{equation}
The temporal low-pass filter for the central signal is $E_{\tau \text{C}}$, while horizontal cells use $E_{\tau \text{S}}$ to simulate delayed photoreceptor impulses. This delay helps the spatio-temporal OPL filter detect luminance changes over time, even in uniform areas \cite{wohrer2009virtual}. Our implementation uses an infinite impulse response (IIR) filter, which is more efficient than a finite impulse response (FIR) filter. The current output, $Y(n)$, is computed using previous output values, $Y(n - i)$.

\begin{equation}
    Y(n)=  \sum_{j=0}^{M} b_{j} X(n-j)-  \sum_{i=1}^{N} a_{i} Y(n-i) 
\end{equation}
where $X(n-j)$ is the previous input samples,  $a_{i}$, and $b_{j}$ are the coefficients of the filter. Fig. \ref{lowpass} illustrates the kernel employed for low-pass filtering in the temporal domain.
\paragraph{Temporal high-pass filter}
In the retina, high-pass filtering occurs due to synaptic antagonism between inhibitory and excitatory cells \cite{wohrer2008model}. Horizontal cells in the OPL and amacrine cells in the IPL are inhibitory cells . This response is obtained by removing low-frequency components, from original signal represented by a Dirac function.

\begin{equation}
T_{\omega,\tau}=\delta_\text{0}-\omega E_{\tau}
\end{equation}

The temporal high-pass filter, $T_{\omega,\tau}$, with weight $\omega$, and the low-pass filter, $E_{\tau}$, enable the digital retina to emulate phasic and tonic cell behavior. By adjusting $\omega$, phasic cells (transient, $\omega = 1$) and tonic cells (sustained, $\omega < 1$) are modeled. Tonic cells respond continuously to stimuli, even at low contrast, while phasic cells respond briefly to high contrast. This filtering represents short-term luminescence adaptation in photoreceptors\cite{wohrer2008model,huth2018convis}. Fig. \ref{highpass} shows the high-pass filter kernel.

\begin{table}[ht]
    \centering
        \caption{\centering Pseudo-code to implement OPL layer} \hrule
    \begin{tabular}{|p{0.95\linewidth}| }
     \textbf{OPL layer}: Spatio-temporal OPL layer with center-surround structure and luminance adaptation expressed through the utilization of a high-pass filter. \\ \hline
      \textbf{Required}: Predefined kernels $K1$ with size 3$\times$3 and $K2$ with size 5$\times$5. Input video frame set X with size 128$\times$128$\times$2000, and each frame is indicated by $x$. Coefficients of IIR low-pass filters ($a1,a3,b1,b3$) and IIR high-pass filter ($a2,b2$), Step size 0.001.  
       \\ \hline
           1. \hspace{14mm} def OPLlayer (X, $a1, a2, a3, b1, b2, b3, K1, K2$) \\     
           2. \hspace{20mm} $G_\text{C}$ = conv2D ($x$, $K1$) \\
           3. \hspace{20mm} $E_{\tau \text{C}}$ = iir\_Lpf ($G_\text{C}$, $a1, b1$)\\
           4. \hspace{20mm} $T_{\text{w},\tau}$ = iir\_hpf ($E_{\tau \text{C}}$, $a2, b2$) \\
           5. \hspace{20mm} $G_\text{S}$ = conv2D    ($T_{\text{w},\tau}$, $K2$)  \\ 
           6. \hspace{20mm} $E_{\tau \text{S}}$= iir\_Lpf ($G_\text{S}$, $a3, b3$)\\
           7. \hspace{20mm} $S$ = $E_{\tau \text{S}}$ \\
           8. \hspace{20mm} $C$ = $T_{\text{w},\tau}$   \\
           9. \hspace{20mm}$I_\text{OPL}$  =  $C$ - $S$ \\
           10.\hspace{14mm} end def \\
       \hline
    \end{tabular}
    \label{opl2}
\end{table}

The OPL layer is a spatio-temporal filter with a band-pass response, utilizing models like the difference of Gaussian (DoG) and difference of exponential (DoE) filters. It is a non-separable filter, meaning spatial and temporal components cannot be separated. The pseudo-code in Table \ref{opl2} represents the OPL layer implementation. The input (X) consists of frames with dimensions 128$\times$128$\times$2000. Each frame ($x$) undergoes spatial filtering via 2-dimensional convolution operation conv2D with 3$\times$3 kernel ($K1$). Temporal low-pass filtering (iir\_Lpf) and high-pass filtering (iir\_hpf) are then applied to each frame, generating the center output signal. Spatial filtering is applied to further process the central output signal using a 5$\times$5\ kernel ($K2$). The resulting signal is delayed through another low-pass filter (iir\_Lpf) to create the surround signal. The center and surround signals are subtracted to generate the resulting signal ($I_\text{OPL}$ current). The spatially linked DoG filter in the OPL exhibits biphasic temporal behavior, serving as both an edge and a motion detector.

  \begin{figure}[!t]%
    \centering
    \subfloat[\centering  ]{{\includegraphics[scale=0.5]{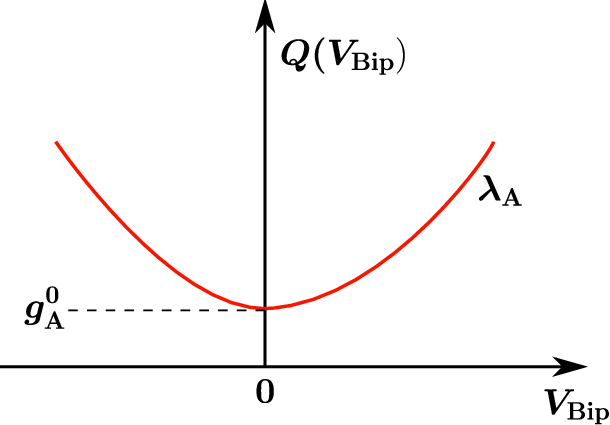} }%
    \label{qfuncmarch03}}
    \qquad    
    \subfloat[\centering]
    {{\includegraphics[scale=0.5]{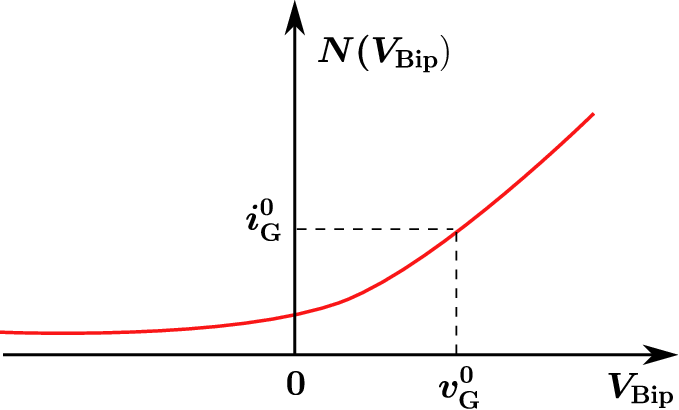} }%
    \label{gi_res}}
    \qquad
   
    \caption{\protect\subref{qfuncmarch03} Shows the activation function, denoted as $Q(V_\text{Bip})$ for the synaptic conductances $g_{A}$ associated with bipolar level contrast adaptation. This implies that the activation of $g_{A}$ is solely determined by the absolute value of $V_\text{Bip}$, the bipolar potential;  \protect\subref{gi_res} Shows the function that rectifies signals from bipolar to ganglion cells, denoted as $N(V_\text{Bip})$. It is a reflection of static nonlinearities that have been empirically discovered in the retina\cite{chichilnisky2001simple,baccus2002fast,kim2001temporal}. }%
    \label{kernels}%
\end{figure}
\subsection{Bipolar-Level Contrast Gain Control Mechanism}

Contrast gain control refers to the regulation of output gain by input contrast between bipolar and amacrine cells \cite{baccus2002fast,shapley1978effect}. This has been shown experimentally by varying contrast stimuli and measuring input-output responses. The implementation uses leaky integrators, which model neural behavior as low-pass filters \cite{wohrer2008model}. The contrast gain control equation from \cite{huth2018convis,wohrer2009virtual} (Eq. \ref{gain1}, \ref{gain2}, \ref{gain3}) describes how bipolar cells integrate the linear OPL current, with feedback from shunt conductances regulating the potential $V_\text{Bip}$.
\begin{equation}
     C\frac{dV_\text{Bip}}{dt} = I_\text{OPL} - g_\text{A}V_\text{Bip}
     \label{gain1}
 \end{equation}
\begin{equation}
       g_\text{A} = G_\text{A}*E_\text{A}*Q(V_\text{Bip})
      \label{gain2}
 \end{equation}
  \begin{equation}
      Q(V_\text{Bip}) = g^0_\text{A} + \lambda_\text{A}V^2_\text{Bip}
      \label{gain3}
 \end{equation}
\begin{table}[ht]
    \centering
        \caption{\centering Pseudo-code for the bipolar-level contrast gain control}
 \hrule
    \begin{tabular}{|p{0.95\linewidth}| }
     \textbf{Contrast gain control}: Contrast adaptation among bipolar and amacrine cells \\ \hline
     \textbf{Required}: Predefined kernel $K3$ with size 5$\times$5. Coefficients of IIR low-pass filter ($a5, b5$) and $g^0_\text{A}$.
       \\ \hline
         
            1. \hspace{14mm} def Bipolar ($K3$, $g^0_\text{A}$, $a4, b4$)  \\
           2. \hspace{20mm} $g_\text{A}$ = $g^0_\text{A}$ + $prev$\_$E_\text{A}$ \\
           3. \hspace{20mm} $att\_map$ = exp (-steps × $g_\text{A}$)  \\
           4. \hspace{20mm} $E_{\infty}$ = ($inputamp$ × $I_\text{OPL}$)    \\
           5. \hspace{20mm} $V_\text{Bip}$ = (($prev$\_$V_\text{Bip}$ - $E_{\infty}$) × \\
          \hspace{30mm} $att\_map$) + $E_{\infty}$ \\ 
           6. \hspace{20mm} $E_\text{A}$ =  (($prev$\_$V_\text{Bip}$)$^2$ × $b4$ \\
            \hspace{36mm} - $prev$\_$E_\text{A}$ × $a4$)\\
           7. \hspace{20mm} $E_\text{A}$ = conv2D ($E_\text{A}$, $K3$)   \\
           8. \hspace{20mm} $prev$\_$V_\text{Bip}$ = $V_\text{Bip}$   \\
           9. \hspace{20mm} $prev\_att\_map$ = $att\_map$ \\
           10.\hspace{19.5mm} $prev$\_$E_\text{A}$ = $E_\text{A}$ \\
           11. \hspace{14mm}  end def\\
       \hline
    \end{tabular}

    \label{vb}
\end{table}

The smoothing of synaptic connections between amacrine cells uses filters $E_\text{A}$ for temporal and $G_\text{A}$ for spatial smoothing. The non-linear gain control is represented by the $Q$ function, with the bipolar cell leakage or "shunt" denoted as $g_\text{A}$, influenced by $Q$. In this model, $g_\text{A}$ varies based on recent feedback signals, as described by Eq. (\ref{gain2}) with specific time and spatial parameters. Eq. (\ref{gain3}) introduces $g^0_\text{A}$ for inert leaks during membrane integration, while $\lambda_\text{A}$ defines the gain control feedback loop's strength.

The pseudo-code in Table \ref{vb} outlines the steps for implementing the contrast gain control equation. First, Eq. (\ref{gain1}) is discretized for computational purposes. The input to this stage is the output of the OPL layer, denoted as $I_\text{OPL}$. $I_\text{OPL}$ undergoes filtering using an exponential kernel, $att_map$, with a time constant $g_\text{A}$, resulting in the variable $V_\text{Bip}$. Pseudo-code lines 2, 6, and 7 implement Eq. (\ref{gain2}) and (\ref{gain3}). In line 6, ($prev$\_$V_\text{Bip}$)$^2$, representing $Q(V_\text{Bip})$ from Eq. 
 (\ref{gain3}), is passed through an IIR low-pass filter, resulting in $E_\text{A}$. The next step involves applying spatial filtering to $E_\text{A}$ using a 5$\times$5 kernel ($K3$) through a conv2D operation. The filtered output is combined with a constant leakage value ($g^0_\text{A}$) to create the total conductance $g_\text{A}$, as described in Eq. (\ref{gain2}). Notably, the variations in $g_\text{A}$ influenced by feedback from $V_\text{Bip}$ contribute to the nonlinear contrast gain control properties.
\subsection{The IPL and Ganglion Cells}
Retinal neurons such as photoreceptor cells, horizontal cells, bipolar cells, and amacrine cells are non-spiking. However, the retina sends the information about the scene as spikes. These spikes are generated by the retinal ganglion cells (RGC) and travel via their axons and optic nerve to higher processing areas of the visual cortex. 
\subsubsection{Excitatory synaptic current in the ganglion layer}
The IPL in the biological retina consists of synaptic connections among bipolar cells, amacrine cells, and ganglion cells. In the digital retina, the formula presented in \cite{huth2018convis,wohrer2009virtual} is used to simulate signal shaping from bipolar cells to ganglion cells, specifically for cat X and Y cells, and primate parvo and magnocellular cells.
\begin{equation}
    I_\text{Gang} = N(\xi\hspace{0.1cm}T_{G}*V_\text{Bip})
    \label{gipe}
\end{equation}
\begin{figure*}[h]%
    \centering
      \subfloat[\centering]{{\includegraphics[scale=0.192]{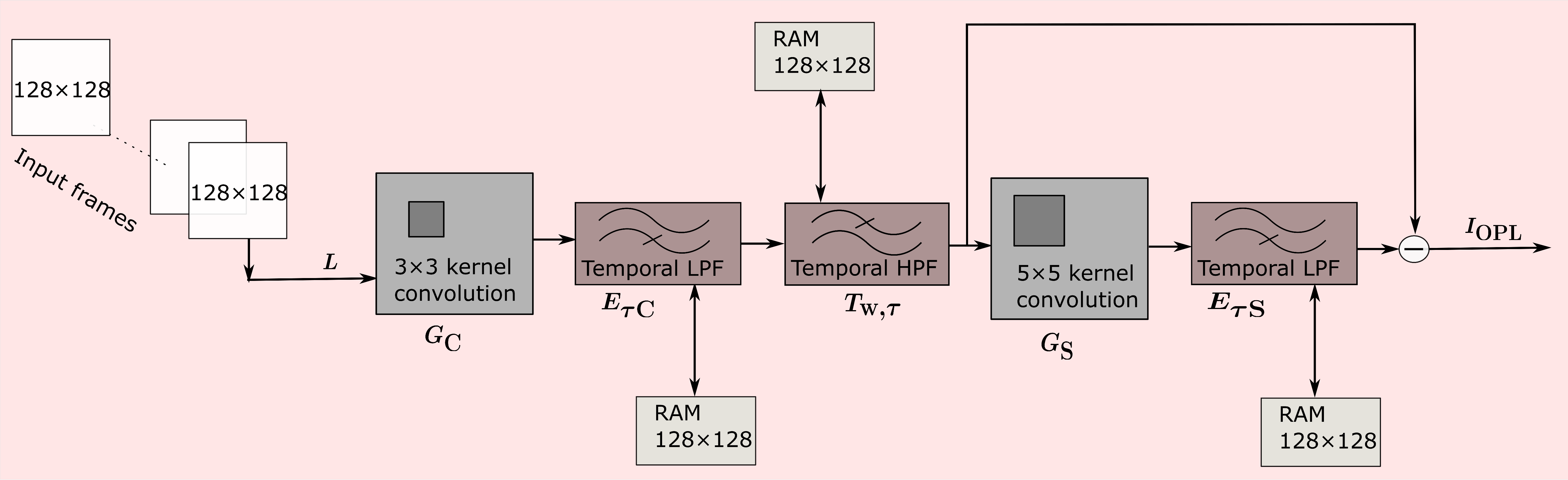}}%
    \label{ooplarch}}
    
    \subfloat[\centering]
    {{\includegraphics[scale=0.192]{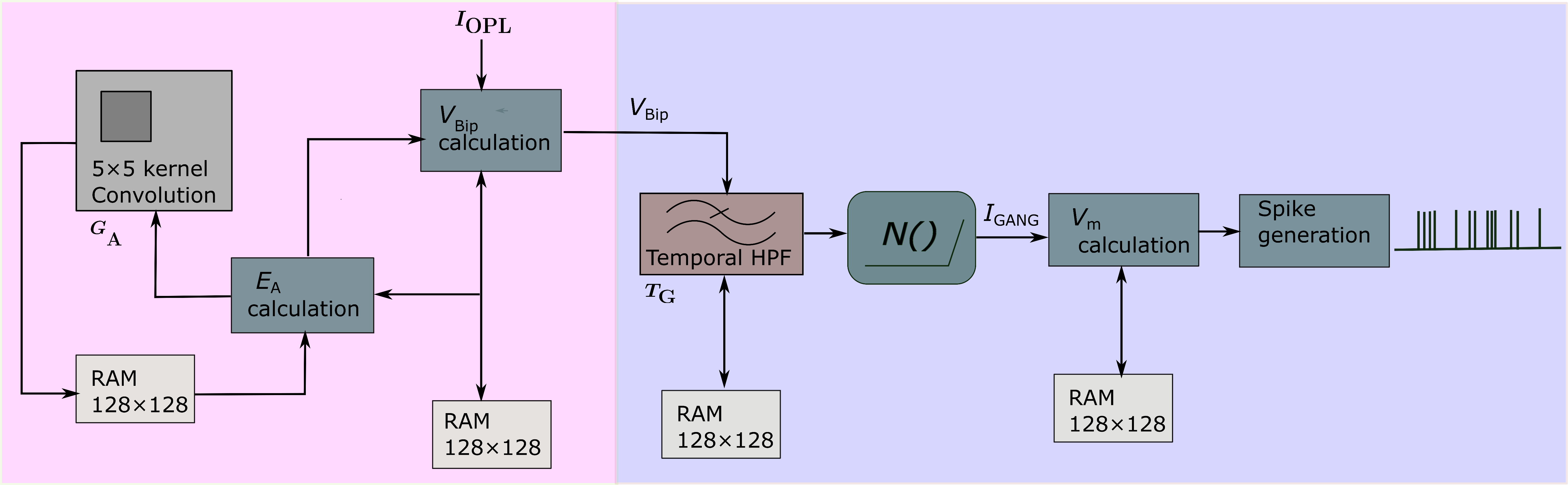}}%
    \label{bipoarchi}}
    \qquad
   \caption{The digital architecture of \protect\subref{ooplarch} the outer plexiform layer (OPL) accepts video frames of size 128$\times$128 as input for its photoreceptor cells and generates corresponding output frames called $I_\text{OPL}$, also with a size of 128$\times$128; \protect\subref{bipoarchi} Depicts that the pink-shaded region represents a bipolar layer incorporating contrast gain control. It accepts $I_\text{OPL}$ as the input frame with a size of 128$\times$128, from the OPL layer and generates an output frame of the same size, 128$\times$128, known as $V_\text{Bip}$ which is gain controlled. Excitatory current $I_\text{Gang}$ is generated within the blue-shaded area and serves as the input to a LIF neuron responsible for spike generation.}%
    \label{systemarch}%
\end{figure*}

\begin{figure*}[!b]
    \centering
       \subfloat[]{{\includegraphics[width=0.92\linewidth, center]{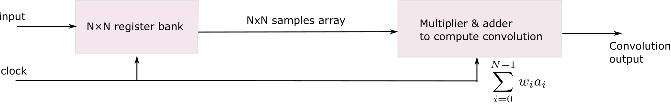} }%
    \label{convblock}}  
    
    \subfloat[]{\includegraphics[width=0.48\linewidth]{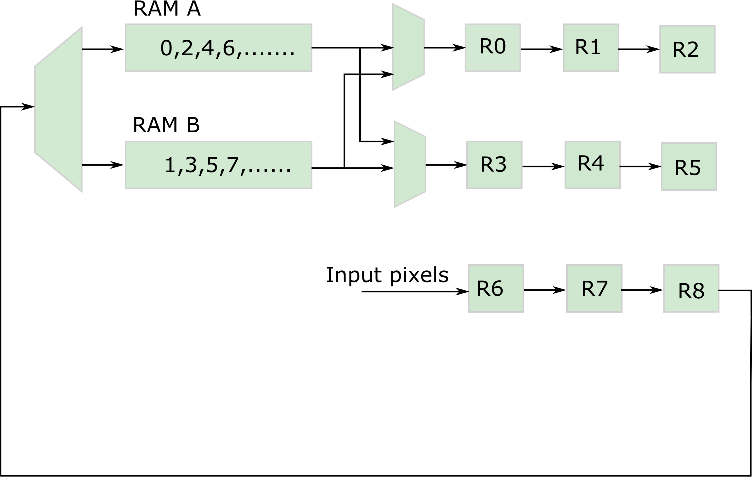}
    \label{conv33}}
    \subfloat[]{\includegraphics[width=0.5\linewidth]{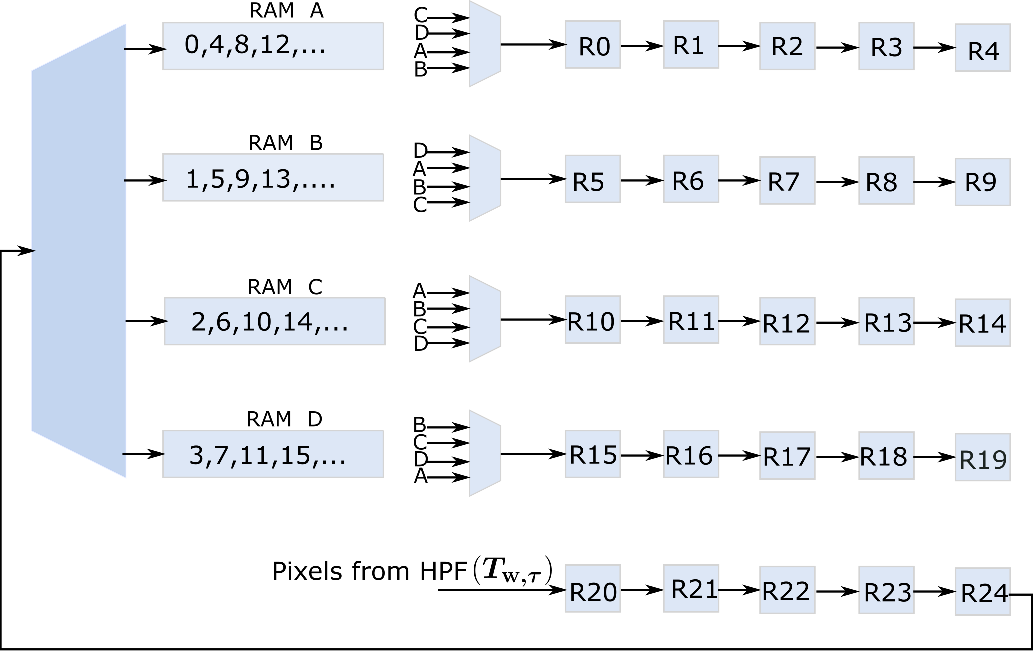}
    \label{conv55}}
      \caption{\protect\subref{convblock} Shows the block diagram of convolution operation for spatial filtering in retinal layers. \protect\subref{conv33} Shows the architecture of $3\times3$ register bank for spatial filtering in the center signal components of the OPL layer. \protect\subref{conv55} Shows the architecture of $5\times5$ register bank for spatial filtering in the surround signal components of the OPL layer, and bipolar level contrast gain control stage.}%
\end{figure*}

where $T_{G}$ is temporal high-pass filter and $N(V_\text{Bip})$ is static nonlinearity function as depicted in Fig. \ref{gi_res}.\\ 
The static nonlinearity function $N(V)$ is given by,
 $$  N(V) = \frac{\textit{i}^0_\text{G}}{1-\lambda_\text{G}(Vthe -\textit{v}^0_\text{G})/\textit{i}^0_\text{G}}\quad  if V \textless  \textit{v}^0_\text{G}$$
$$N(V) =\textit{i}^0_\text{G} + \lambda_\text{G}(V-\textit{v}^0_\text{G}) \quad if V \textgreater  \textit{v}^0_\text{G}$$
where parameter ${v}^0_{G}$ is the threshold for the linearity of the cell with its corresponding value ${i}^0_\text{G}$, transmission then turns linear after that. The scaling parameter of the nonlinearity function is denoted as $\lambda_\text{G}$. Pseudo-code in Table \ref{ig2} models the ganglion input layer. This layer receives the gain-controlled signal $V_{Bip}$ and filters it using the high-pass filter iir\_hpf. Logical indexing is applied in line 5 to remove negative samples from the filtered data. Lines 3, 4, and 5 implement the static non-linearity function. Generated spikes can be categorized as ON spikes or OFF spikes based on the value of variable $\xi$. If $\xi$ is set to 1, the spikes are ON spikes, and if $\xi$ is -1, they are OFF spikes.
\begin{table}[ht]
    \centering
        \caption{\centering Pseudo-code for ganglion input layer}
  \hrule
    \begin{tabular}{|p{0.95\linewidth}| }
     \textbf{Ganglion cell current}: Excitatory current upon ganglion cell. \\ \hline
      \textbf{Required}: Coefficients of IIR high-pass filter (iir\_hpf) $a5, b5$, bipolar amplification $\lambda_\text{G}$, parameters of static nonlinarity $\textit{i}^0_\text{G}$ and $\textit{v}^0_\text{G}$.
       \\ \hline
         
           1. \hspace{8mm} def ganglion current ($a5, b5$, $\lambda_\text{G}$, $\textit{i}^0_\text{G}$, $\textit{v}^0_\text{G}$)   \\
           2. \hspace{13mm}   $x$ = $\xi$ × iir\_hpf ($V_\text{Bip}$, $a5, b5$)  \\
           3. \hspace{13mm}  n = ($\textit{i}^0_\text{G}$ × $\textit{i}^0_\text{G}$) / ($\textit{i}^0_\text{G}$ - $\lambda_\text{G}$ × ($x$ - $\textit{v}^0_\text{G}$)),     if $ V < \textit{v}^0_\text{G} $  \\
           4. \hspace{13mm} ngreater = $\textit{i}^0_\text{G}$ + $\lambda_\text{G}$ × ($x$  - $\textit{v}^0_\text{G}$),    if $V > \textit{v}^0_\text{G} $ \\
           5. \hspace{13mm}  n($x$  $>$ 0) = ngreater($x$  $>$ 0) \\ 
           6. \hspace{13mm}  $I_\text{Gang}$ = n \\
           7. \hspace{8mm}  end def\\
       \hline
    \end{tabular}

    \label{ig2}
\end{table}
\subsubsection{Ganglion layer}
\begin{table}[ht]
    \centering
        \caption{\centering Pseudo-code for ganglion output layer}
\hrule
    \begin{tabular}{|p{0.95\linewidth}| }
     \textbf{Spike generation}: The spiking output of the ganglion cells is generated using LIF neurons. \\ \hline
      \textbf{Required}: Leak current $gL$, length of time steps $tau$, refractory time $rt$.
       \\ \hline
         
           1. \hspace{8mm} def spiking ($gL$, $tau$, $rt$)  \\
           2. \hspace{13mm} $V_{m} = (V_{m} + (I_\text{Gang}- gL \times  V_{m})  tau)$ \\
           3. \hspace{13mm} $rt$ = $rt$ - 1  \\
           4. \hspace{13mm} $V_{m}$($rt$ $>=$ 0.5) = zeros($rt$ $>=$ 0.5)  \\
           5. \hspace{13mm} Spikes = $V_{m}$ $>$ 1  \\ 
           6. \hspace{13mm} $rt$($rt$ $<$ 0.0) = zeros($rt$ $<$ 0.0) \\
           7. \hspace{13mm} $V_{m}$(Spikes) = zeros(Spikes)   \\
           8. \hspace{13mm} allspikes = cat(3, allspikes, Spikes)  \\
           9. \hspace{8mm} end def \\
       \hline
    \end{tabular}
     \label{spikes}
\end{table}   
  \begin{table}[]
    \centering
        \caption{\centering The top module of pseudo-codes}
\hrule
    \begin{tabular}{|p{0.95\linewidth}| }
        \textbf{Required}: Frames X, filter coefficients $a1$-$a5$ and $b1$-$b5$, Kernels $K1$-$K3$, $g^0_\text{A}$, $\lambda_\text{G}$, $gL$, $tau$, $rt$.
       \\ \hline
         
           1. \hspace{8mm} for image x in X do \\    
           2. \hspace{13mm} OPL (X, $a1, a2, a3, b1, b2, b3, K1, K2$)   \\
           3. \hspace{13mm} Bipolar ($K3$, $g^0_\text{A}$, $a4, b4$)  \\
           4. \hspace{13mm}  ganglion current ($a5, b5$, $\lambda_\text{G}$, $\textit{i}^0_\text{G}$, $\textit{v}^0_\text{G}$)  \\
           5. \hspace{13mm} spiking ($gL$, $tau$, $rt$) \\ 
           6. \hspace{8mm} end for\\
       \hline
    \end{tabular}

    \label{totallab}
\end{table}

The highly non-linear process of converting rectified current $I_\text{Gang}$ into spike trains involves a leaky integrate-and-fire (LIF) neuron, which generates spikes when input surpasses a threshold, with no output for weak inputs. This LIF model, resembling retinal ganglion cells, converts the continuous signal $I_\text{Gang}$ into spike trains as given by,
\begin{equation}
     C\frac{dV_\text{m}{}}{dt} = I_\text{Gang} - g^{L}V_\text{m}
     \label{giout}
 \end{equation}
 where $V_{m}$ is the membrane potential and $g^{L}$ is the time constant. During the refractory period of the neuron, $V_\text{m}$ remains fixed at 0. Once $V_\text{m}$ reaches a specific threshold, it is then reset to 0.
The digital implementation of the equation is presented in pseudo-code in Table \ref{spikes}. Line 2 describes the discretization of Eq. (\ref{giout}) using the rectified signal $I_\text{Gang}$ as the input for spike generation. The variables $gL$ and $tau$ represent the leak current and the duration of each time step, respectively. The refractory time is denoted by $rt$. Spikes are generated when the membrane potential $V_\text{m}$ exceeds the threshold voltage of 1. Parameters for the implementation of the primate retina \cite{huth2018convis} are listed in Table \ref{tab1}.

\section{Digital Implementation}
\label{fpgaimplementation}
\begin{figure*}[!t]
  \centering
  \includegraphics[scale=0.180]{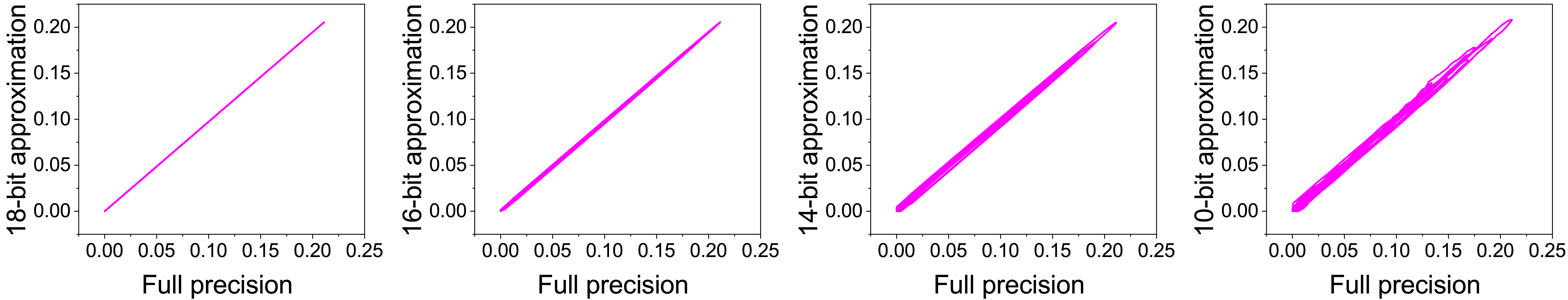}
  \caption{Impact of quantization and approximation of ganglion layer signal ($I_\text{Gang}$).}
  \label{fixed}
\end{figure*}

The mathematical model of the retina is demonstrated using the pseudo-codes shown in Table \ref{opl2}, \ref{vb}, \ref{ig2}, \ref{spikes}, and \ref{totallab}.  
The proposed VLSI architecture is the direct hardware implementation of this mathematical model. The result of decreasing the bit precision on the approximation of the ganglion layer signal $I_\text{Gang}$ is observed in Fig. \ref{fixed}.

\subsection{ VLSI architecture of the OPL layer}

\begin{table}[]
\centering
\caption{FPGA Implementation Summary.}
\begin{tabular}{|cc|}
\hline
\multicolumn{2}{|c|}{\cellcolor[HTML]{C0C0C0}\begin{tabular}[c]{@{}c@{}}\textbf{FPGA Implementation Summary} \\ \textbf{Device Xa7a75tcsg324-2L}\end{tabular}} \\ \hline
\multicolumn{1}{|c|}{Max Clock Frequency (MHz)} & 83 \\ \hline
\multicolumn{1}{|c|}{Slice LUTs} & 3720 \\ \hline
\multicolumn{1}{|c|}{Slice Registers} & 3711 \\ \hline
\multicolumn{1}{|c|}{LUTRAM} & 39 \\ \hline
\multicolumn{1}{|c|}{Slice} & 1720 \\ \hline
\multicolumn{1}{|c|}{Block RAM(36Kb)} & 98.5 \\ \hline
\multicolumn{1}{|c|}{DSPs} & 76 \\ \hline
\multicolumn{1}{|c|}{IO} & 199 \\ \hline
\multicolumn{1}{|c|}{Latency (clock cycles)} & 413 \\ \hline
\multicolumn{1}{|c|}{Throughput (ksps)} & 12 \\ \hline
\end{tabular}
\label{resource}
\end{table}
\begin{figure}[h]%
    \centering
    \subfloat[\centering]{{{\includegraphics[scale=0.172]{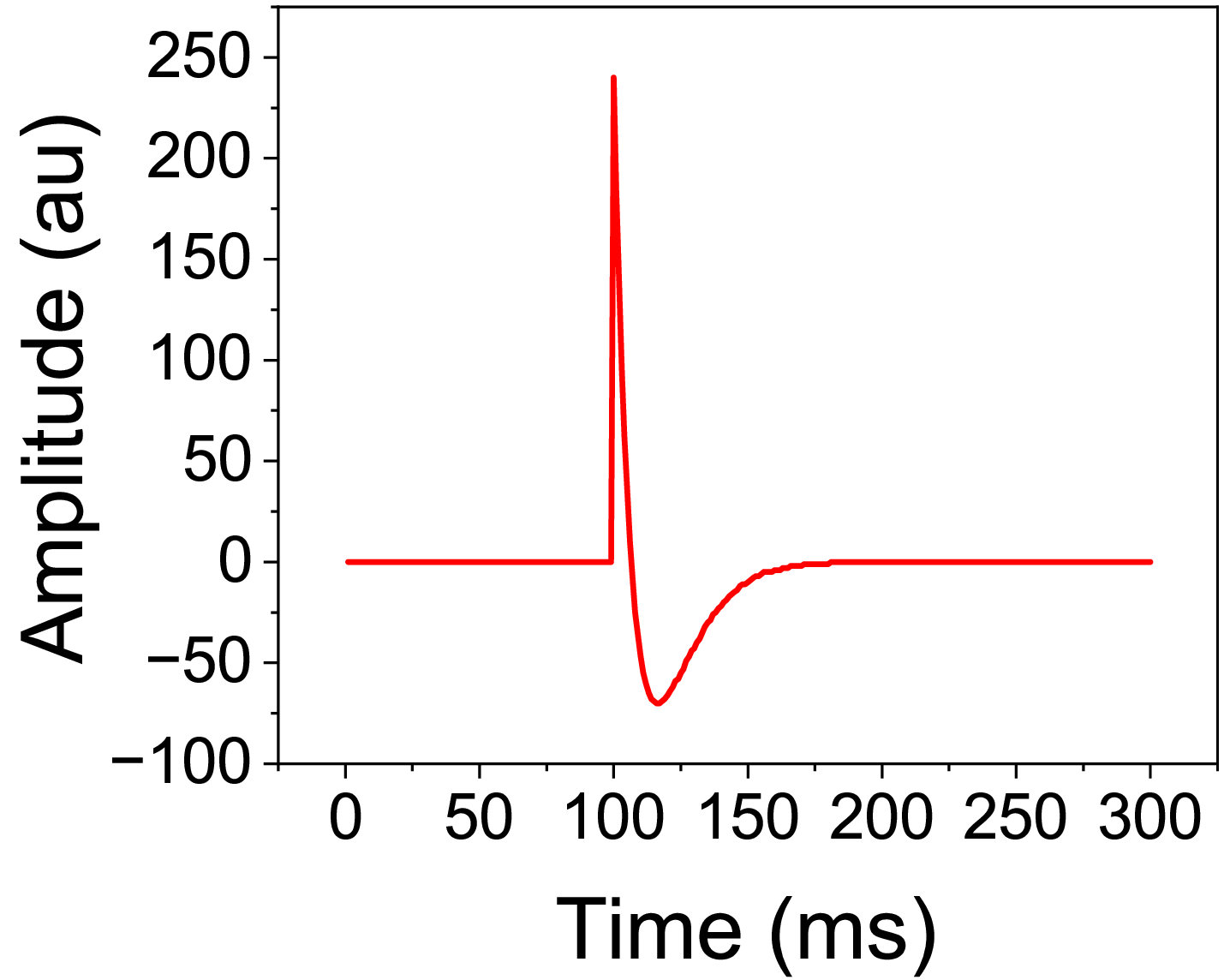}} }%
    \label{psf}}
\subfloat[\centering]{{{\includegraphics[scale=0.18]{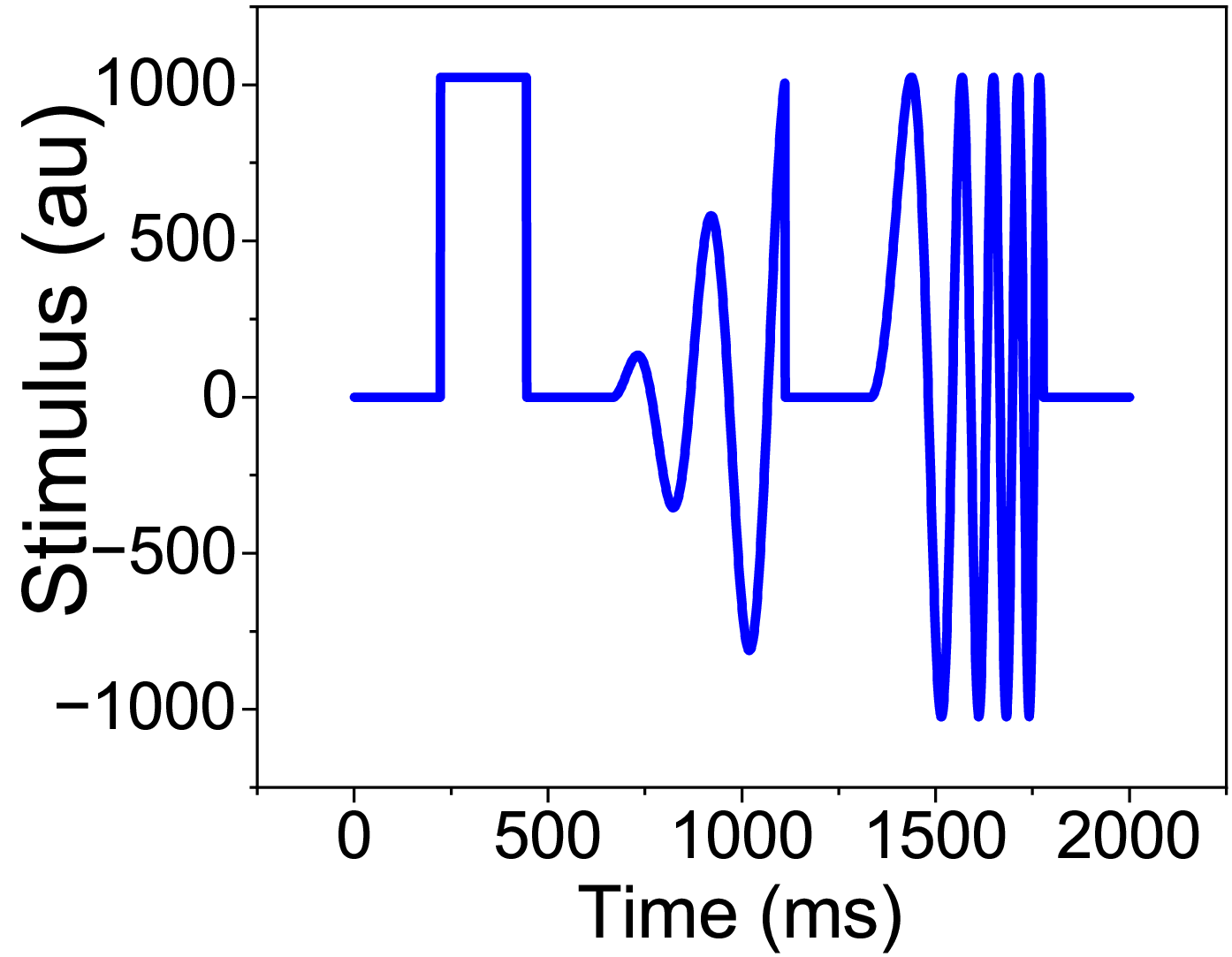}} }%
    \label{chirpstimulus}}
    \qquad
   \caption{\protect\subref{psf} 
   Obtained impulse response of the photoreceptor cell (center signal) for an impulse input applied to the photoreceptor cell of the digital retina, similar to observations by Schnapf \cite{schnapf1990visual} in the macaque retina ('au' denotes arbitrary units);   \protect\subref{chirpstimulus} An oscillatory 'chirp' stimulus, consisting of an OFF-ON-OFF pulse followed by increasing frequency and amplitude oscillations, is used as the luminance input for comparison simulations shown in Fig. \ref{oplchirp}.}%
\end{figure}
 The system-level architecture of the OPL layer depicted in Fig. \ref{ooplarch}, is designed to support a frame size of 128 $\times$ 128 and a frame rate of 200 frames per second (fps). During each clock cycle, it retrieves the luminance value ($L$) of a pixel from a video frame. The initial operation is spatial filtering, applying a Gaussian filter to the input frame through convolution with a 3$\times$3 window size. The general architecture for the convolution operation is shown in Fig. \ref{convblock}. It has two parts: register bank and multiplier and adder unit. The dimension of the register bank is the dimension of the convolution window. The internal architecture and necessary interconnections for the register bank of window size  3$\times$3 are shown in Fig. \ref{conv33}. The nine registers denoted as $R_\text{0} - R_\text{8}$, are used to store luminance values of the current 3$\times$3 window. The multiplier and adder structure is used to compute multiply and addition operations and generate convolution output.

The input pixels are retrieved from a frame in a raster scan manner by the design. Every clock cycle, one input pixel is retrieved and passed through three registers $R_{6}$, $R_{7}$, and $R_{8}$ in three clock cycles and stored in either the RAM A or RAM B. The dimension of each ram is 128$\times$8-bit. Pixels from even-indexed rows are stored in RAM A, while those from odd-indexed rows are stored in RAM B. Specifically, the $0^{th}$ row pixels are stored in RAM A, and the $1^{st}$ row pixels are in RAM B. The convolution operation begins when the first pixel of the $1^{st}$ row is available in $R_{7}$ and the first pixel of the $0^{th}$ row is also available in $R_{4}$ from RAM A. At that moment the content of the other registers $R_{0}$ - $R_{2}$, $R_{5}$, and $R_{8}$ are kept zero to handle zero padding at the image boundary. The register bank is designed in such a way that it can select zero values at the image boundaries based on the control signal generated from the control unit. 

In this design, dual-port RAMs are used to allow for simultaneous reading and writing operation in RAM A and RAM B. The RAMs work in a ping-pong manner. The pipelined implementation of the design improves timing and generates an output at every clock cycle. As a new row is fetched, it replaces the old values of a RAM when they have already been processed.  

 \begin{figure*}[t]
	\centering
	\subfloat[]{\includegraphics[width=0.245\linewidth]{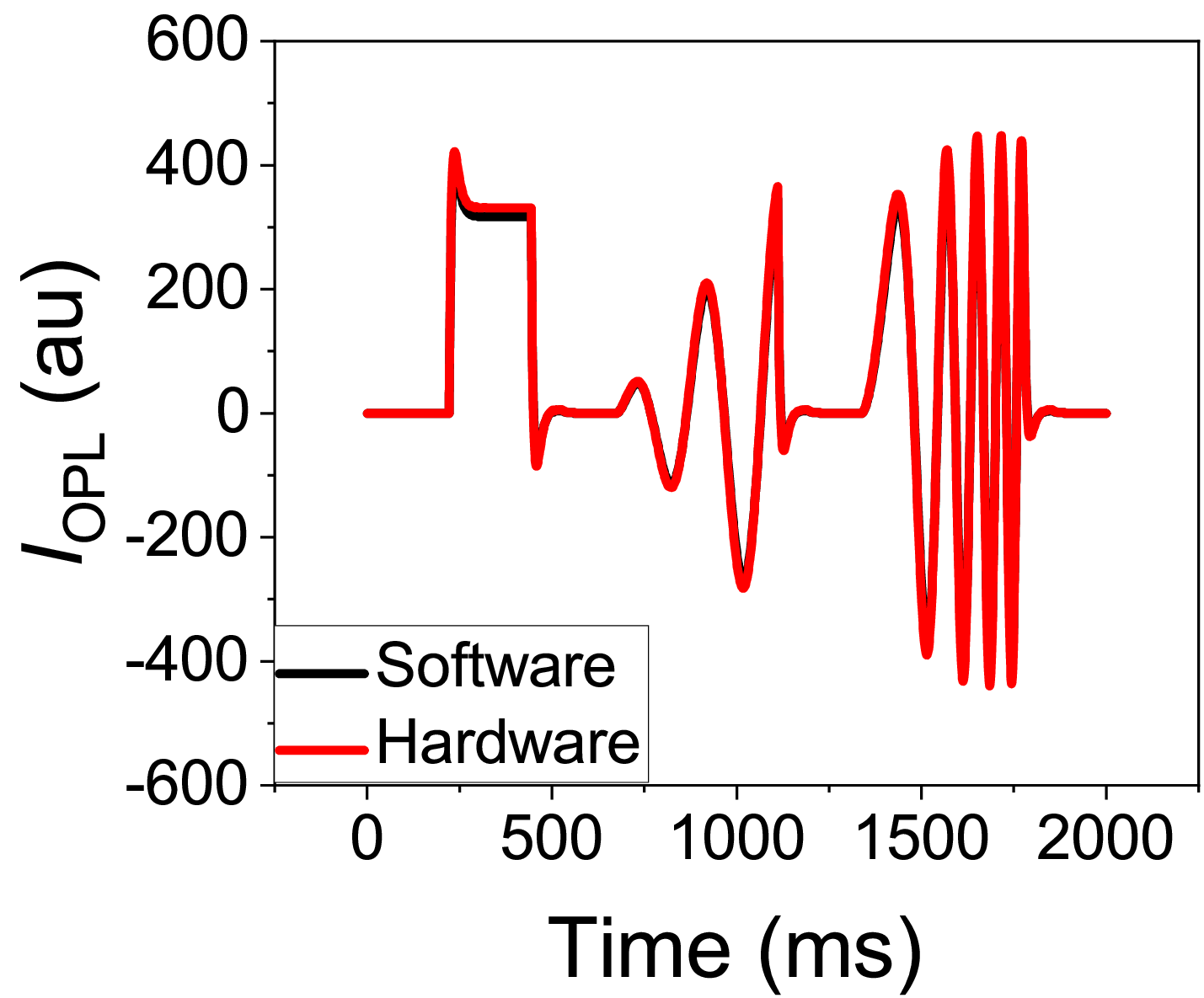}
		\label{opla}}
  \subfloat[]{\includegraphics[width=0.245\linewidth]{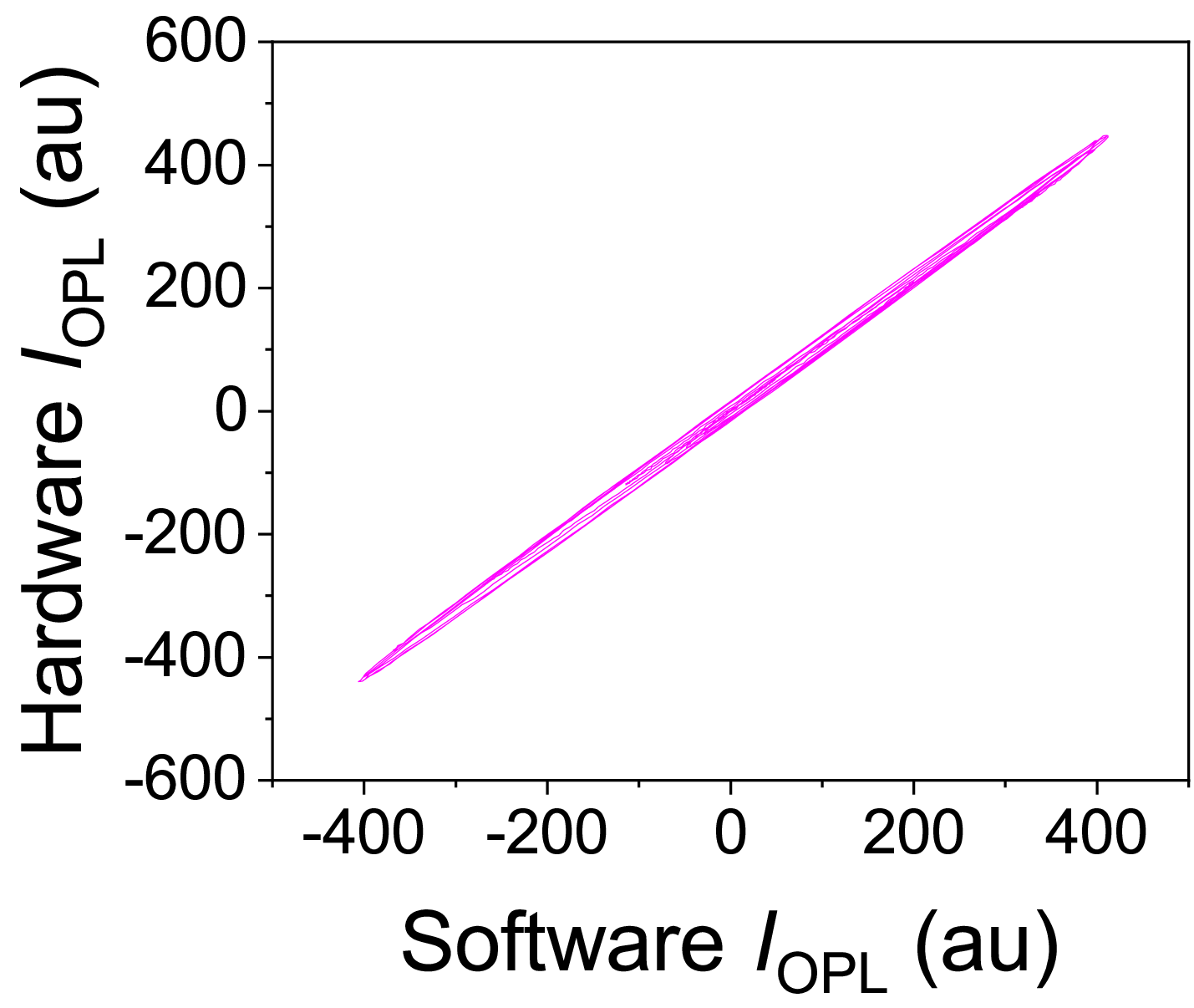}
		\label{oplb}}
     \subfloat[]{\includegraphics[width=0.245\linewidth]{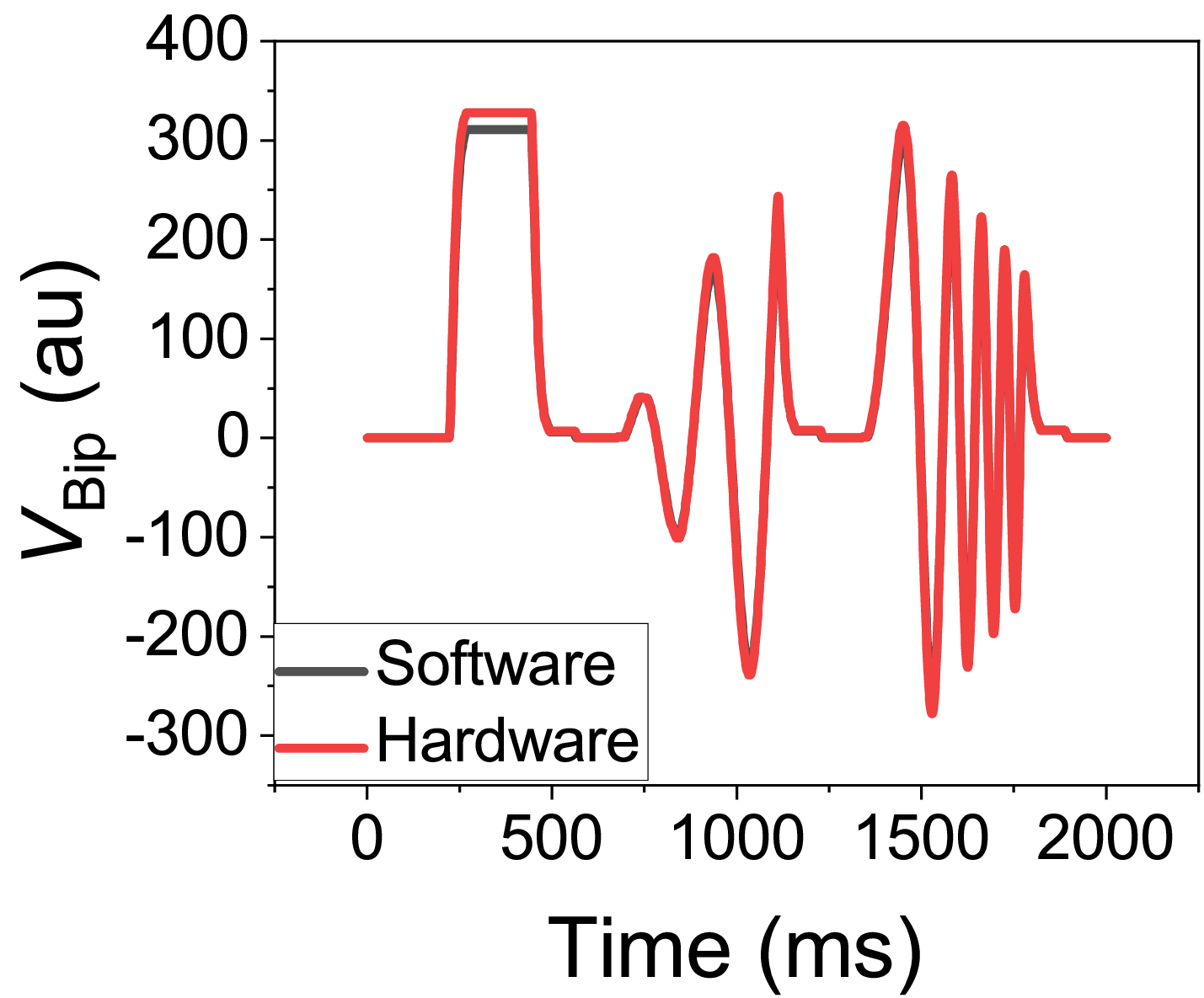}
		\label{bijan}}	
  \subfloat[]{\includegraphics[width=0.25\linewidth]{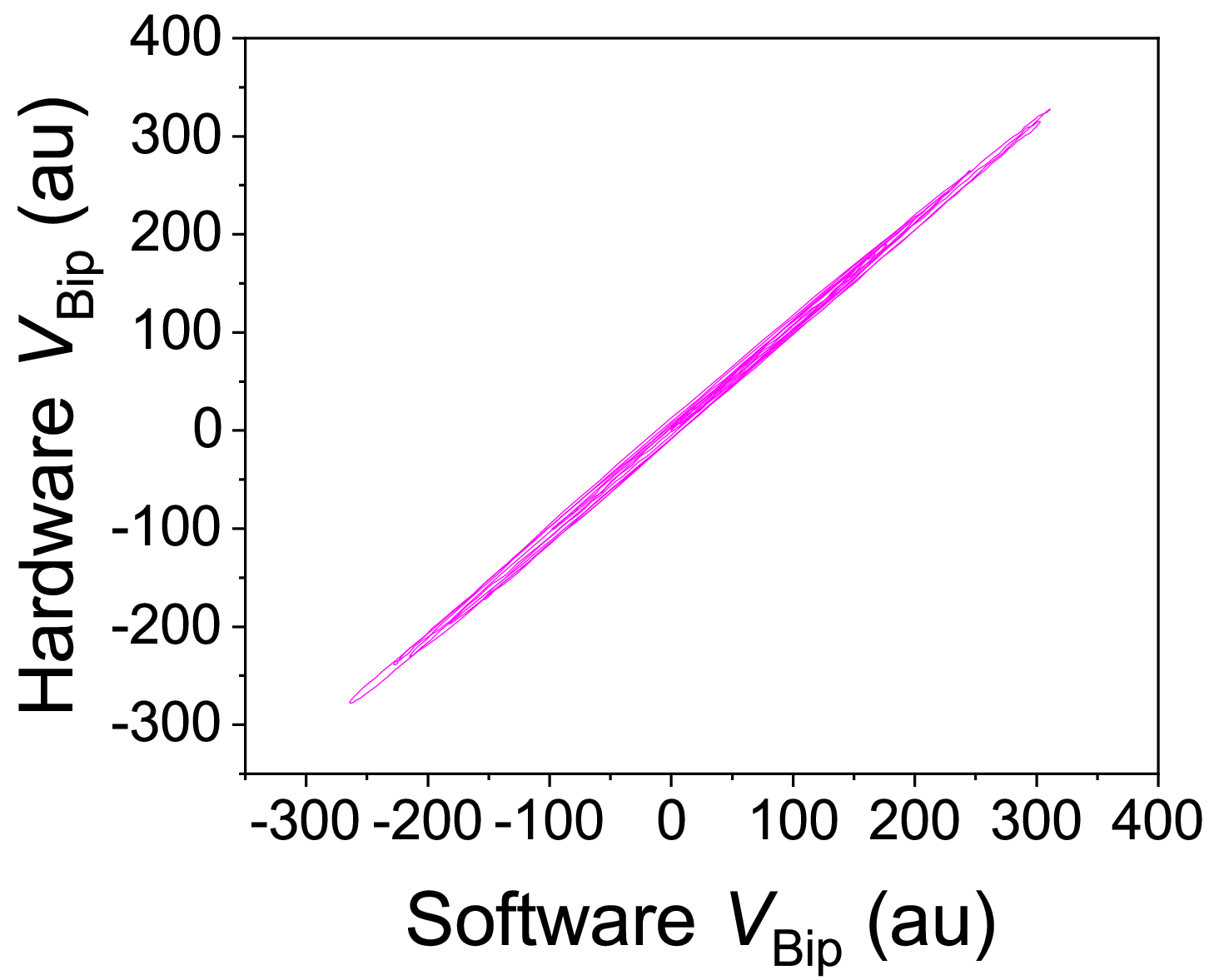}
		\label{bitrace}}\\
  	\subfloat[]{\includegraphics[width=0.245\linewidth]{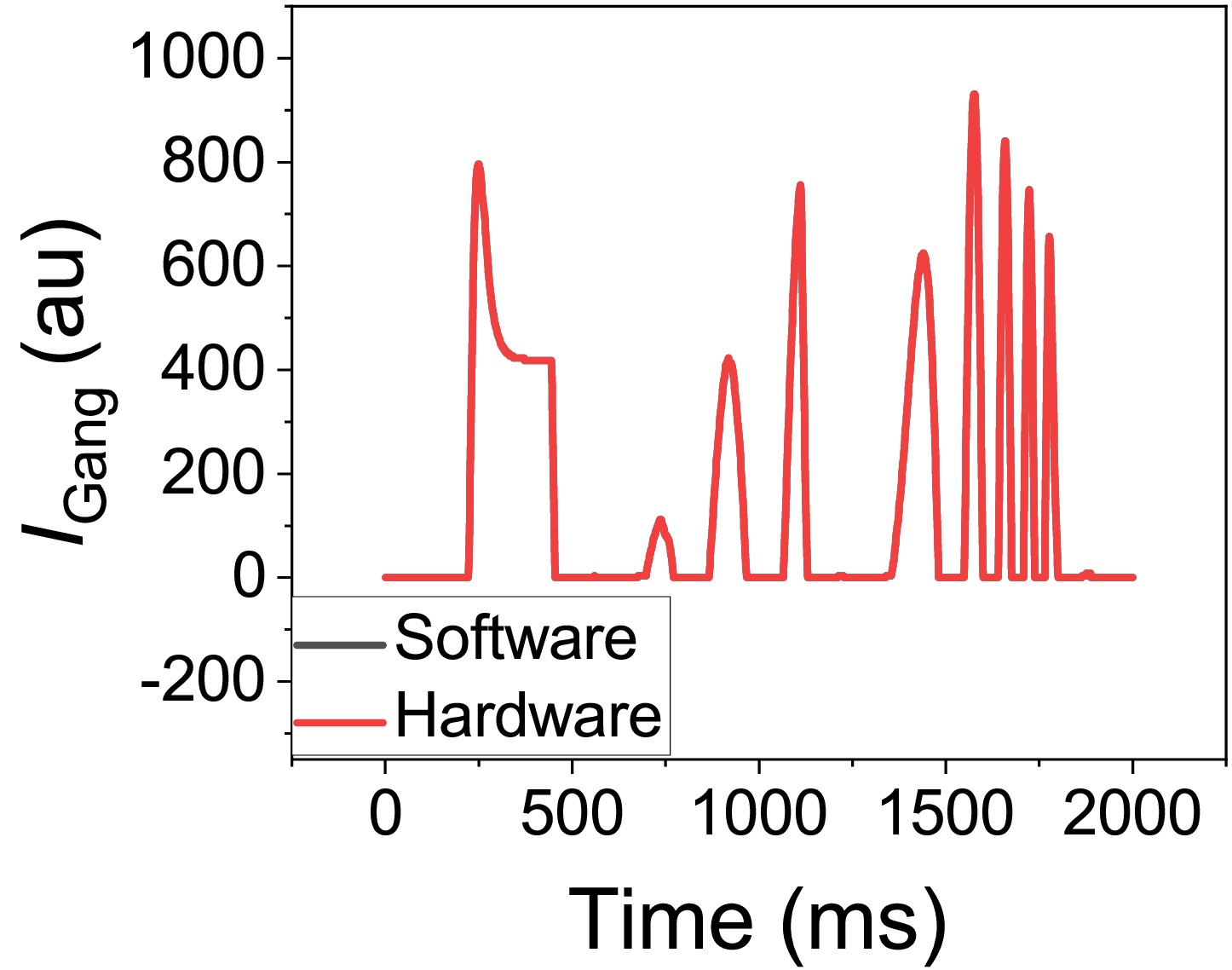}
		\label{gia}}
  \subfloat[]{\includegraphics[width=0.245\linewidth]{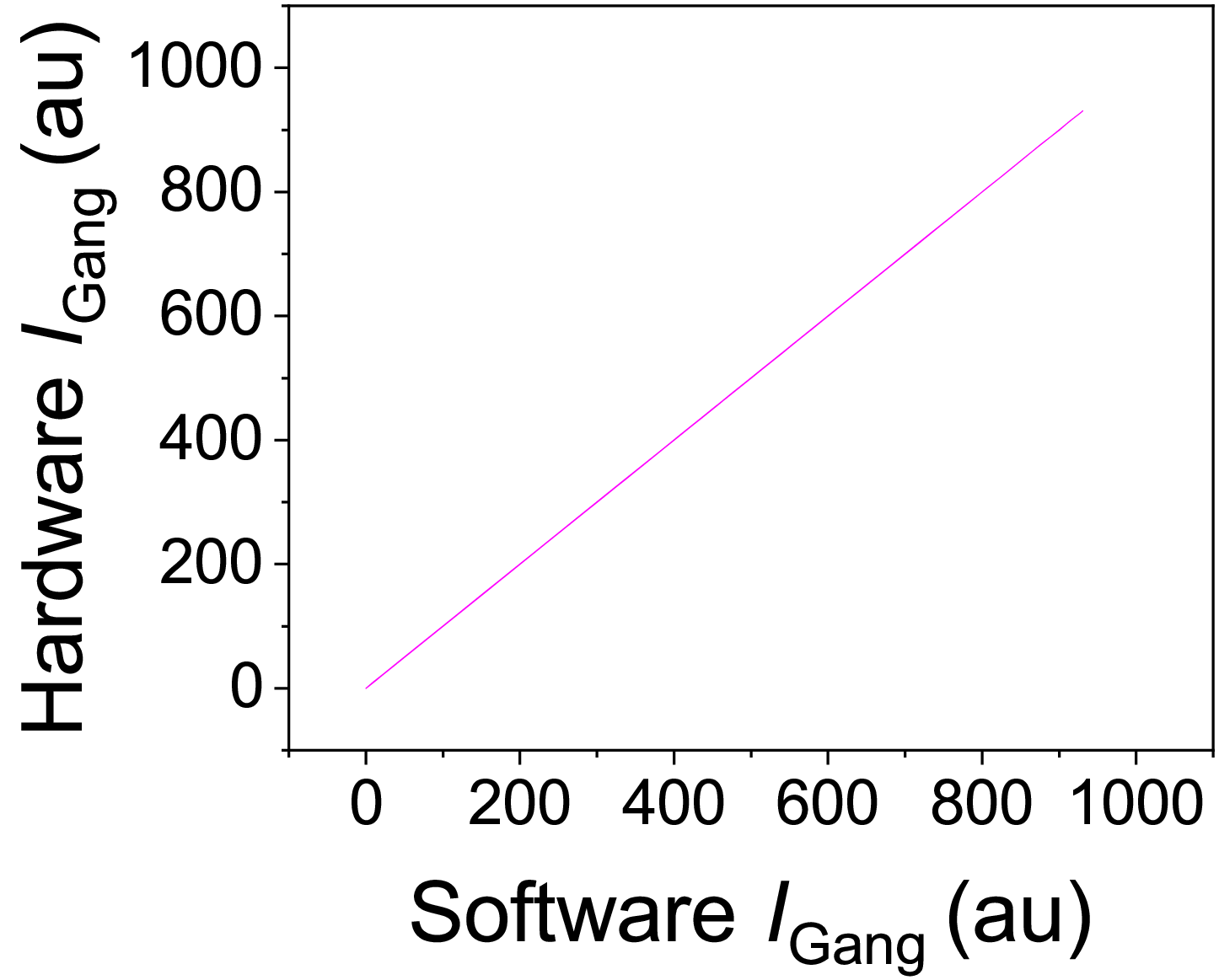}
		\label{gib}}
  \subfloat[]{\includegraphics[width=0.245\linewidth]{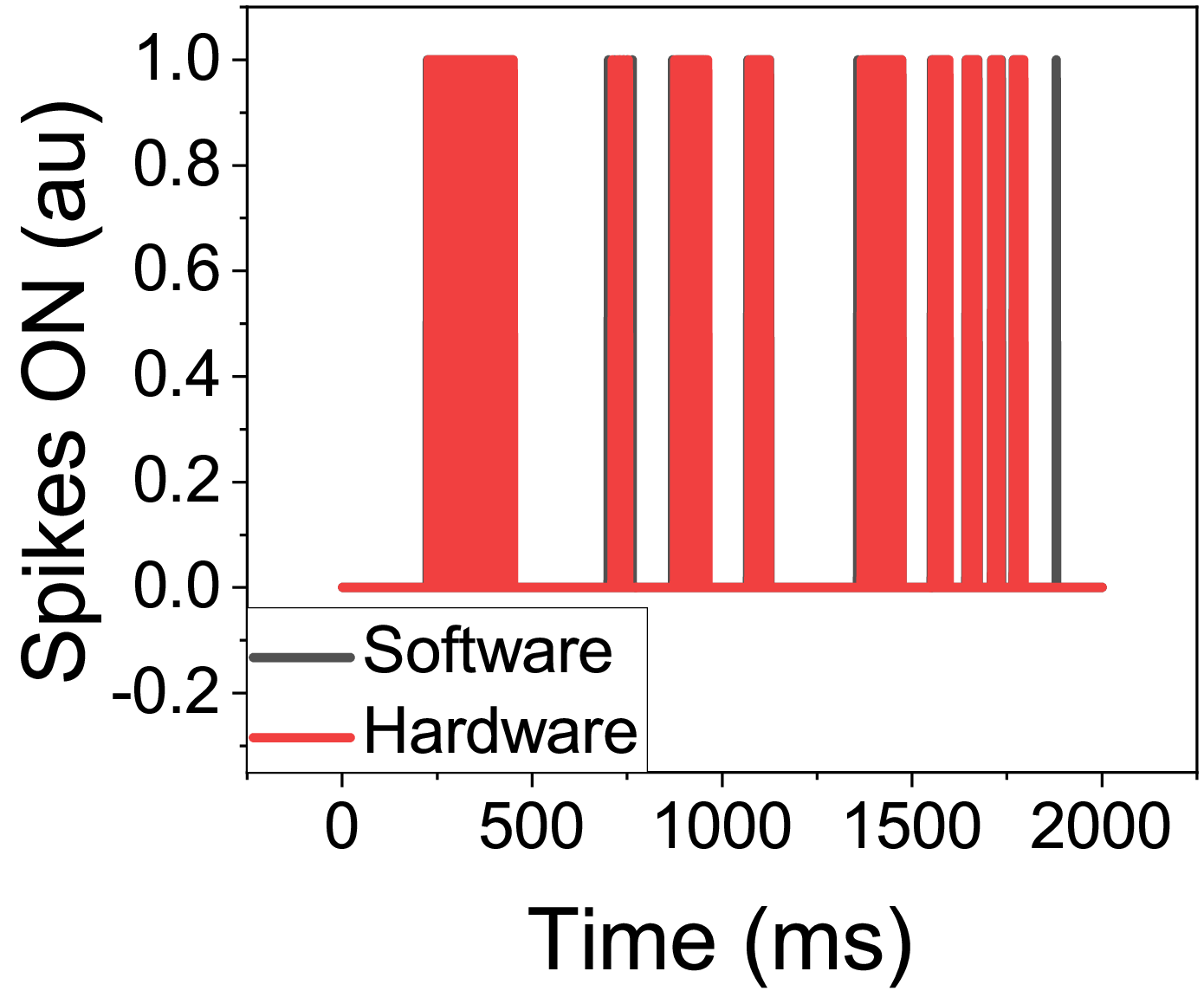}
		\label{spikingA}}
    \subfloat[]{\includegraphics[width=0.245\linewidth]{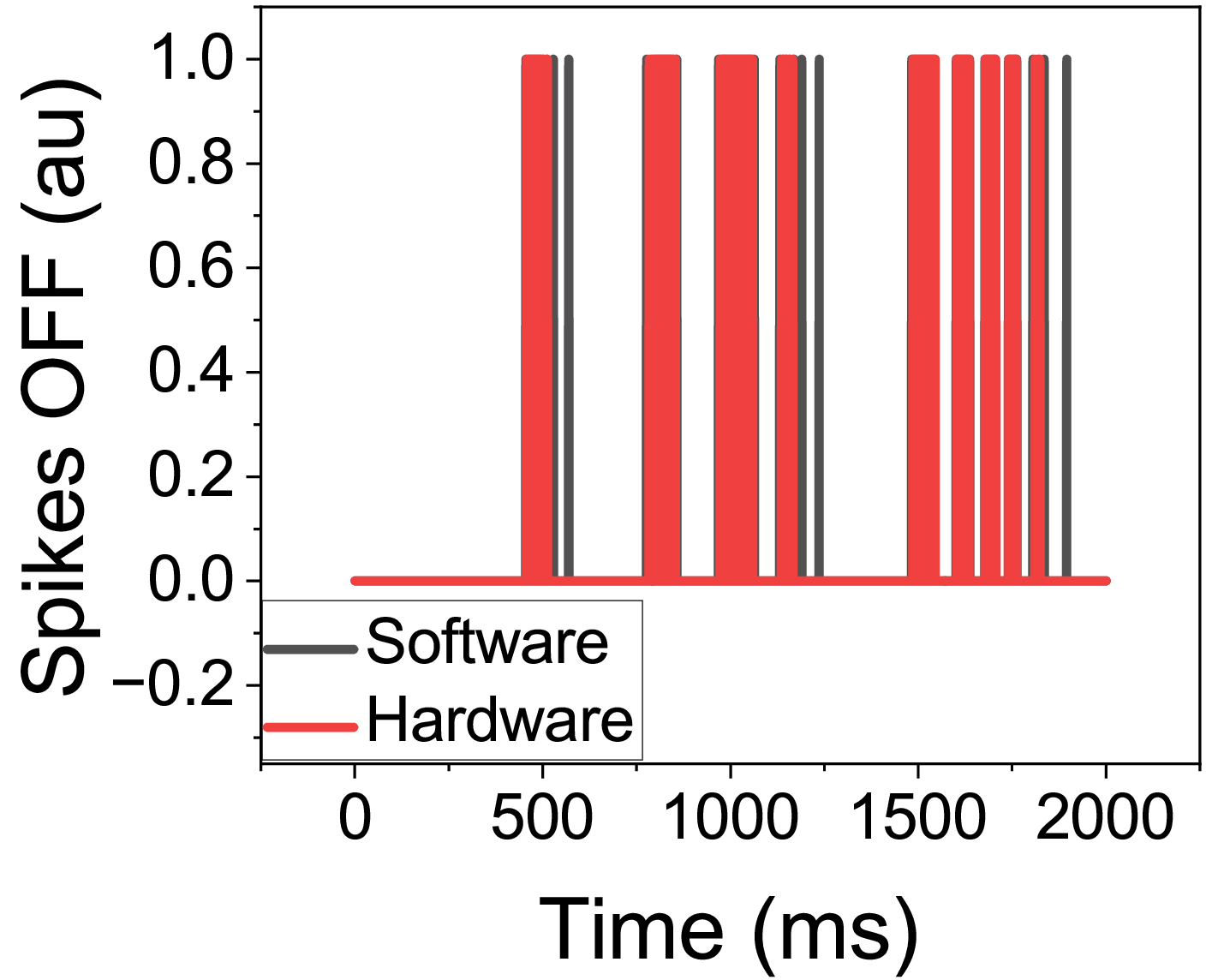}
		\label{spikingOFFA}}
\caption{
\protect\subref{opla} Shows software and digital retina OPL layer signals ($I_\text{OPL}$) to a chirp stimulus (Fig. \ref{chirpstimulus});
\protect\subref{oplb} Compares software and digital retina OPL layer signal traces;
\protect\subref{bijan} Shows software and digital retina bipolar signals ($V_\text{Bip}$);
\protect\subref{bitrace} Compares software and digital retina bipolar signal traces;
\protect\subref{gia} Shows software and digital retina excitatory current on ganglion cells ($I_\text{Gang}$);
\protect\subref{gib} Compares software and digital retina ganglion current traces;
\protect\subref{spikingA} Shows ON spikes from software and digital retina spiking layers to the chirp stimulus (Fig. \ref{chirpstimulus});
\protect\subref{spikingOFFA} Shows OFF spikes from software and digital retina spiking layers to the chirp stimulus (Fig. \ref{chirpstimulus}).}
 \label{oplchirp}
\end{figure*}

\begin{table}[]
\centering
\caption{Parameters of the retina}
\begin{tabular}{|c|c|c|}
\hline
\rowcolor[HTML]{C0C0C0} 
\textbf{Parameters} & \textbf{Floating pt.}&\textbf{Fixed pt.}  \\ \hline
($\sigma_{C}$) & 0.05(deg.) &0.0498\\ \hline
($\tau_{c}$) & 10 ms&9.8 ms\\ \hline
($\tau_{u}$) & 10 ms &9.8 ms\\ \hline
($\sigma_{S}$) & 0.15(deg.)&0.1494 \\ \hline
($\tau_{s}$) & 10 ms&9.8 ms\ \\ \hline
($\lambda_{OPL}$) & 1&1 \\ \hline
($\omega_{OPL}$) & 0.5 &0.5\\ \hline
($\sigma_{A}$) & 0.05(deg.) &0.0498 \\ \hline
($\tau_{A}$) & 5 ms &4.9 ms\\ \hline
($g^0_{A}$) & 50&50 \\ \hline
($\lambda_{A}$) & 0 &0\\ \hline
($\tau_{G}$) & 20 ms&19.5 ms \\ \hline
($\lambda_{G}$) & 5 &5\\ \hline
($\textit{i}^0_\text{G}$)&0.008&0.0078\\ \hline
($\textit{v}^0_\text{G}$)&0&0\\ \hline
($g^{L}$)&0.1&0.0996\\ \hline

\end{tabular}
 \label{tab1}
\end{table}

  \begin{figure}[]%
    \centering
    \subfloat[\centering  ]{{\includegraphics[scale=0.19]{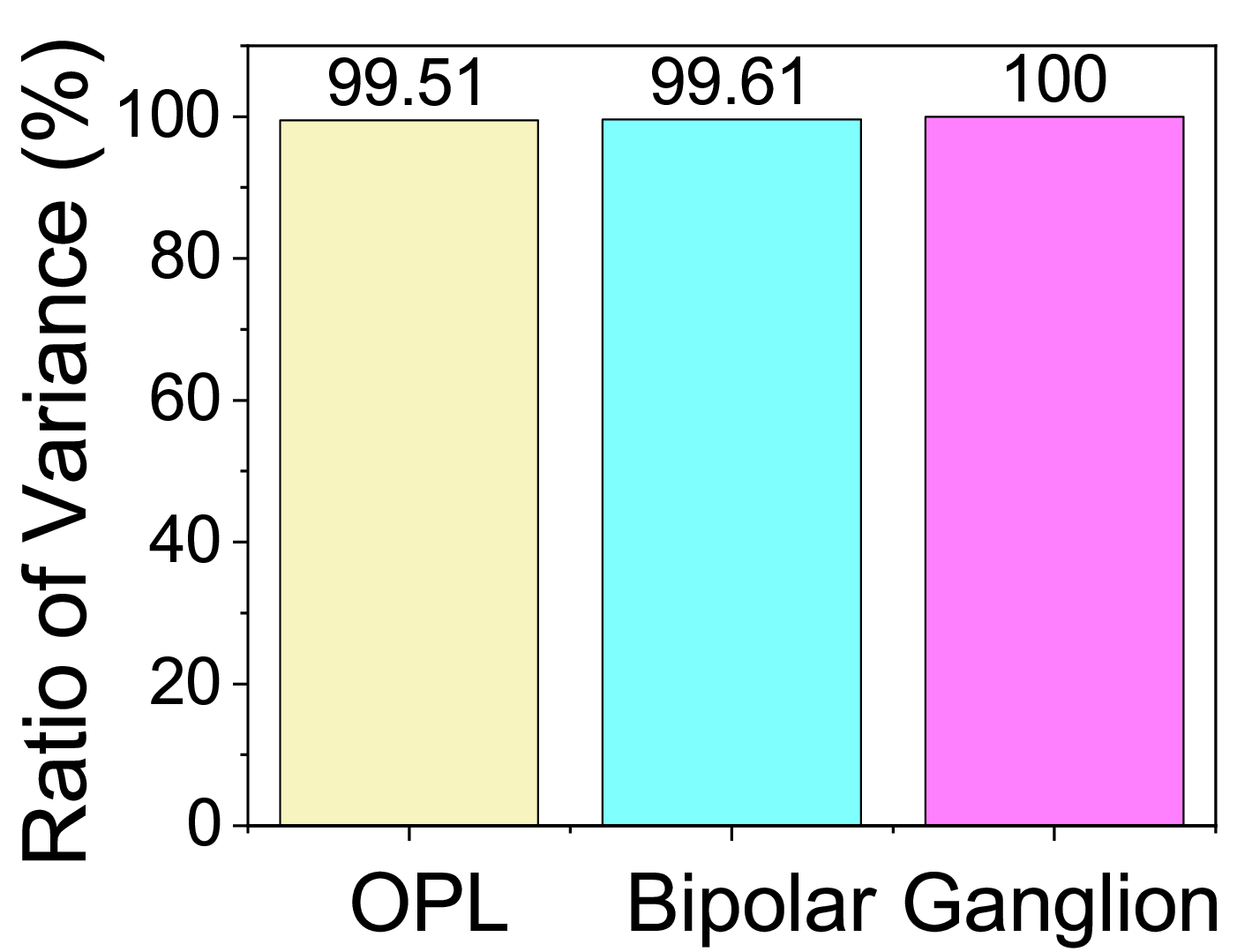} }%
    \label{variance}}        
    \subfloat[\centering]    
    {{\includegraphics[scale=0.18]{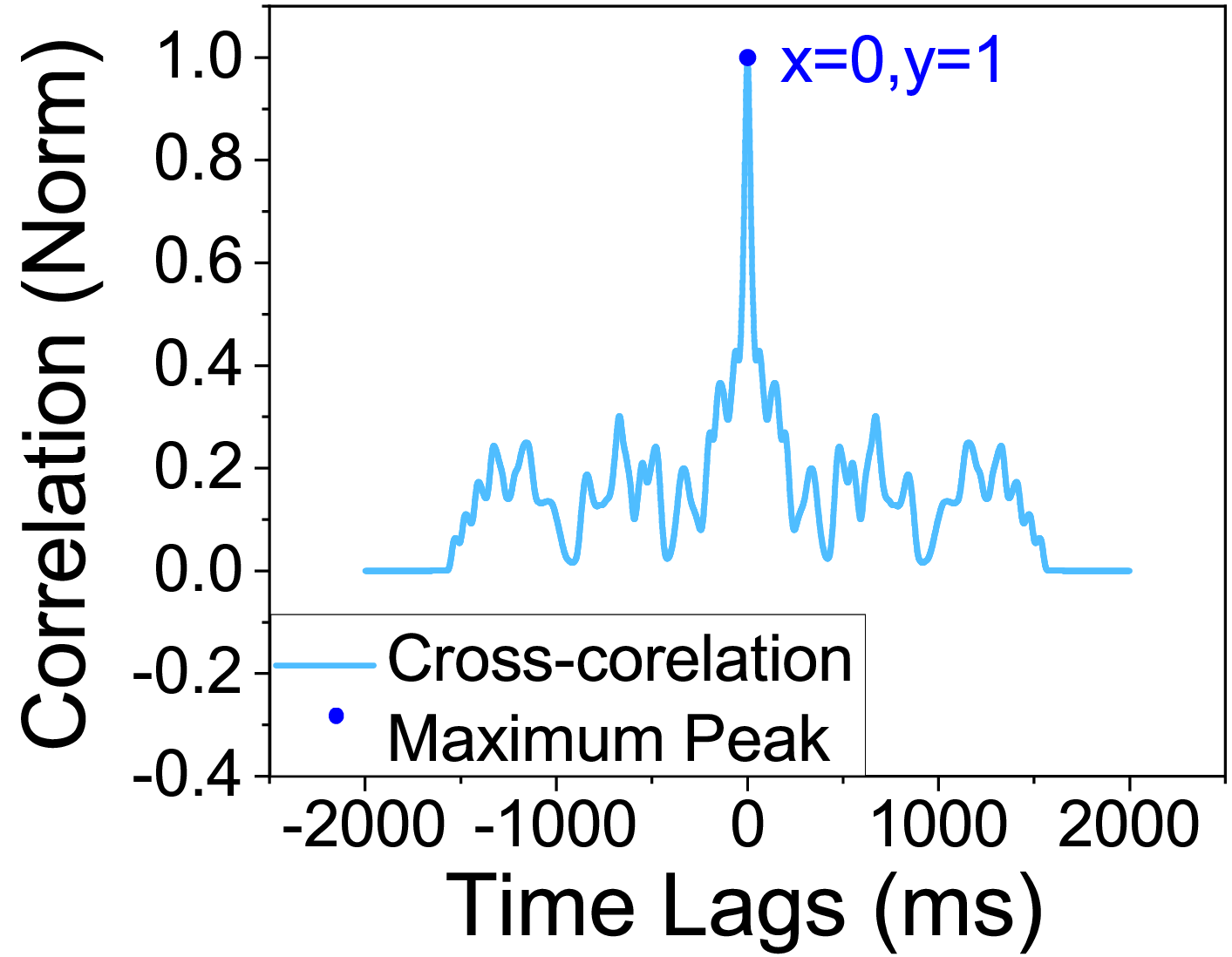} }%
    \label{coor2}}
    \qquad
   \caption{\protect\subref{variance} Shows the percentage of software retina model's variation that is explained for each stage of the digital retina individually in the comparison simulations. Specifically, the OPL layer ($I_\text{OPL}$), bipolar layer ($V_\text{Bip}$), and excitatory current on ganglion cells ($I_\text{Gang}$) are analyzed from Fig. \ref{oplchirp}; \protect\subref{coor2} Shows time difference of the hardware and software retina model.}%
\end{figure}   
  \begin{figure*}[]
        \vspace{-1cm}
	\centering 
	\subfloat[]{\includegraphics[width=0.247\linewidth]{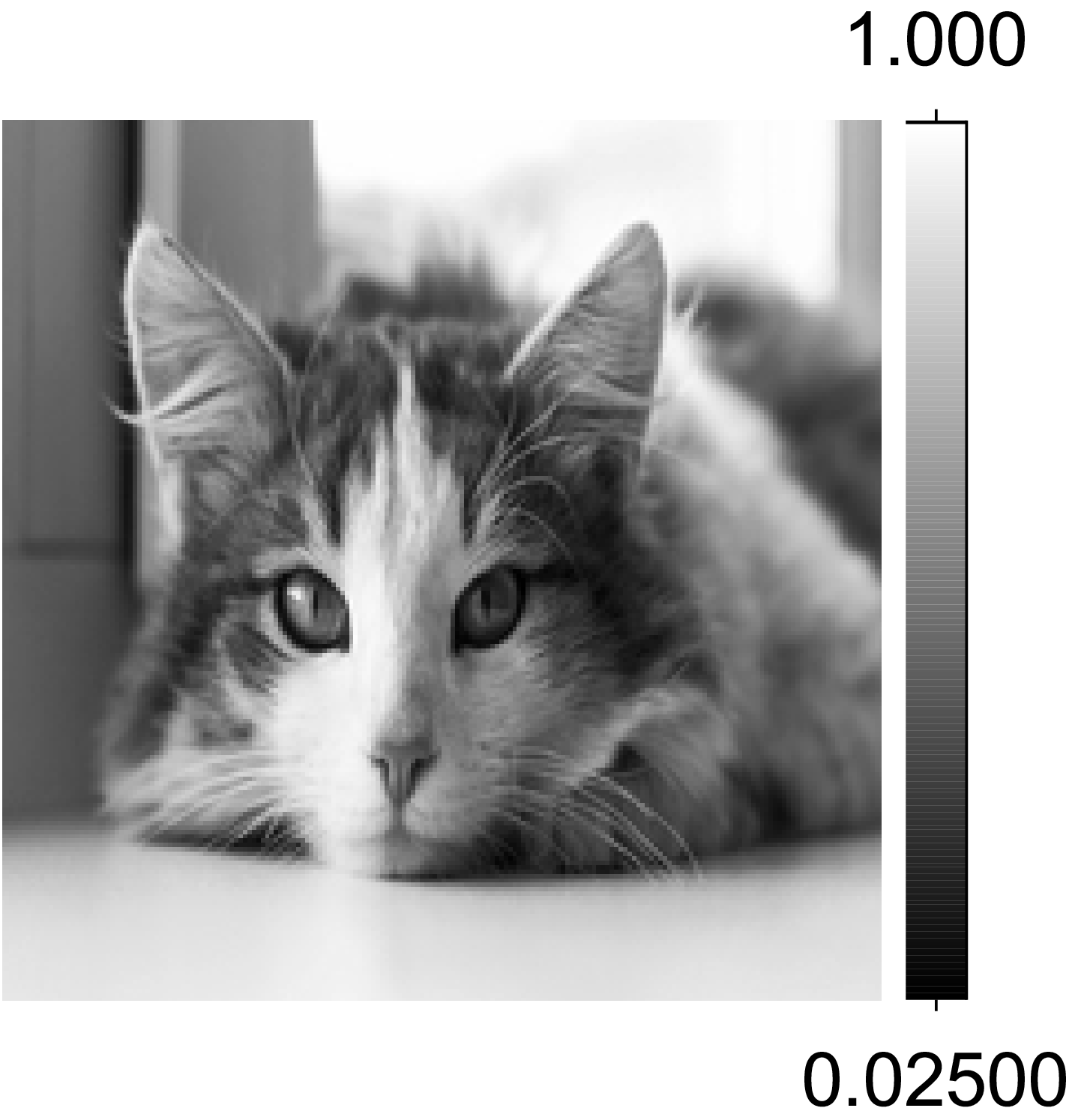}
		\label{imagein}}
  \subfloat[]{\includegraphics[width=0.247\linewidth]{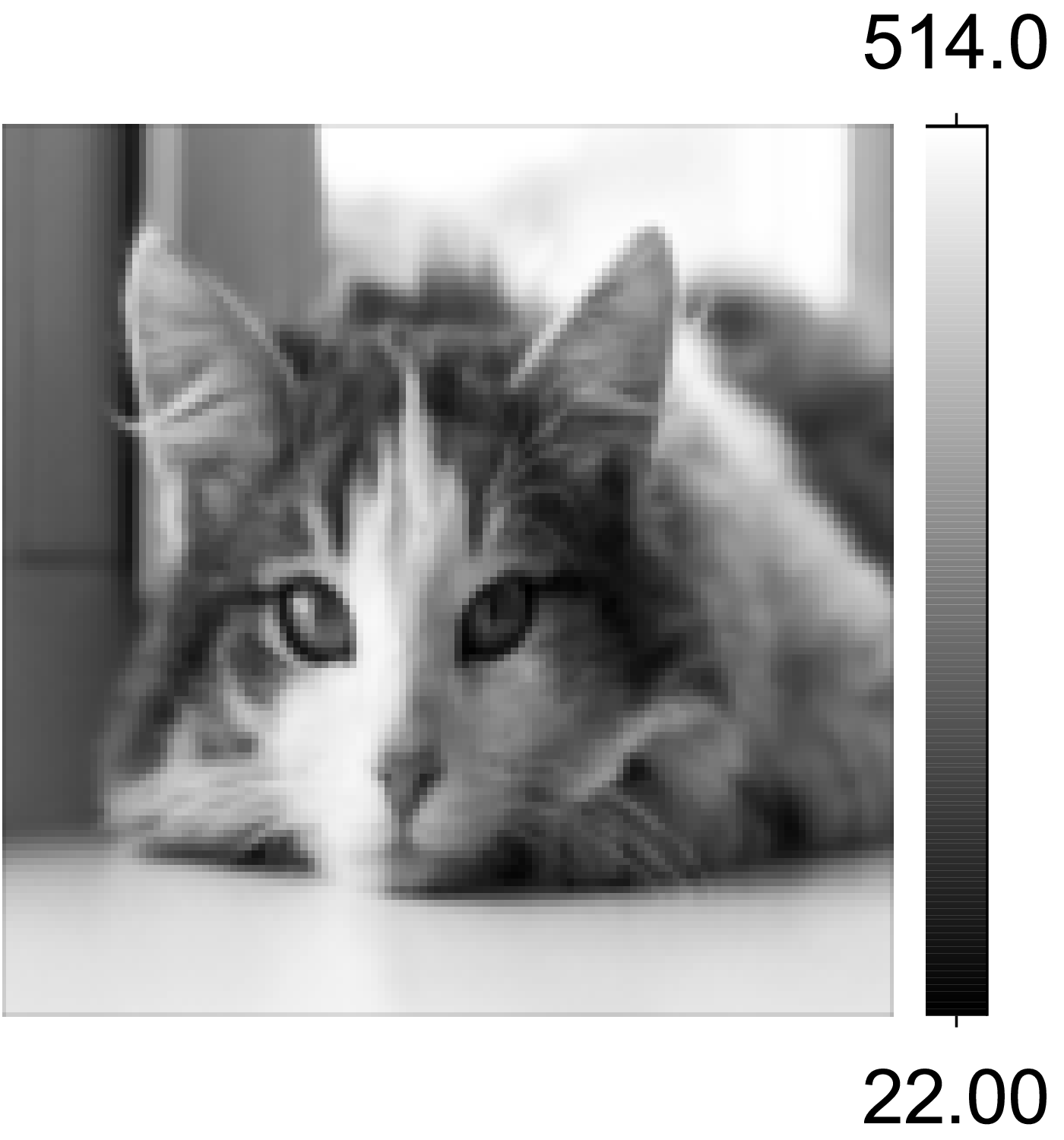}
		\label{centerimage}}
	\subfloat[]{\includegraphics[width=0.247\linewidth]{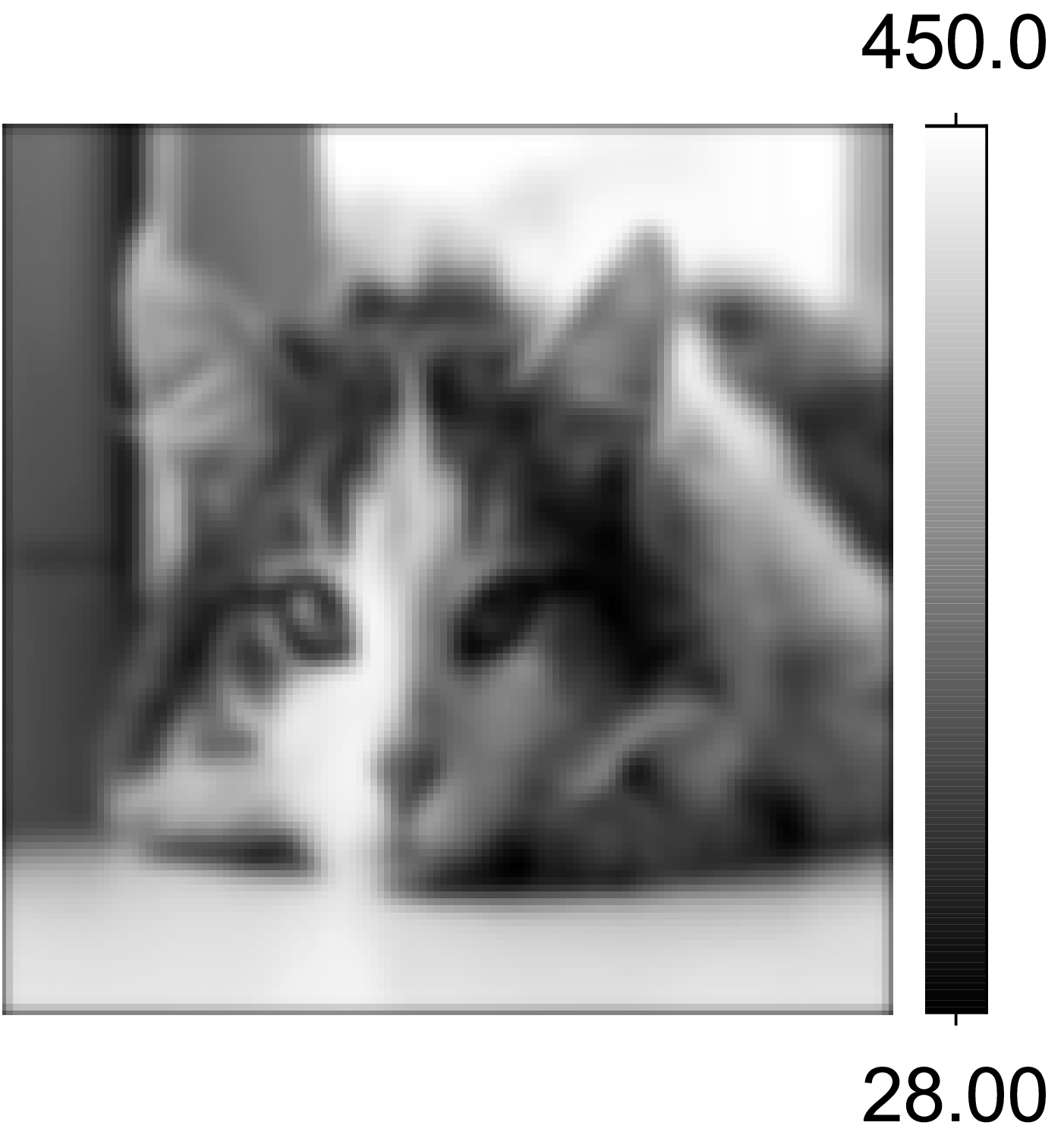}
		\label{surroundimage}}  
  \subfloat[]{\includegraphics[width=0.247\linewidth]{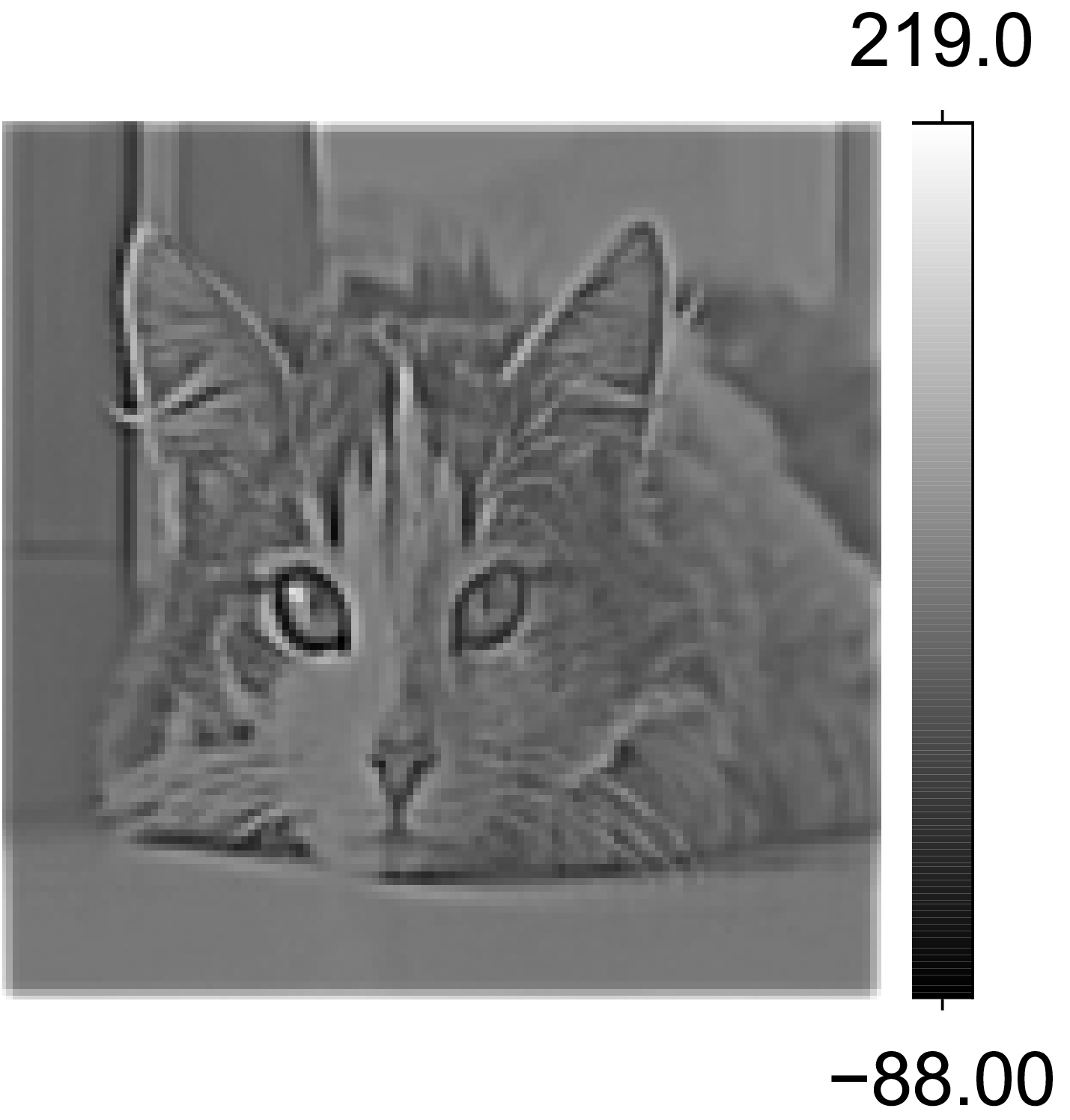}
		\label{opljan}}
    \vspace{-0.50cm}
	\subfloat[]{\includegraphics[width=0.247\linewidth]{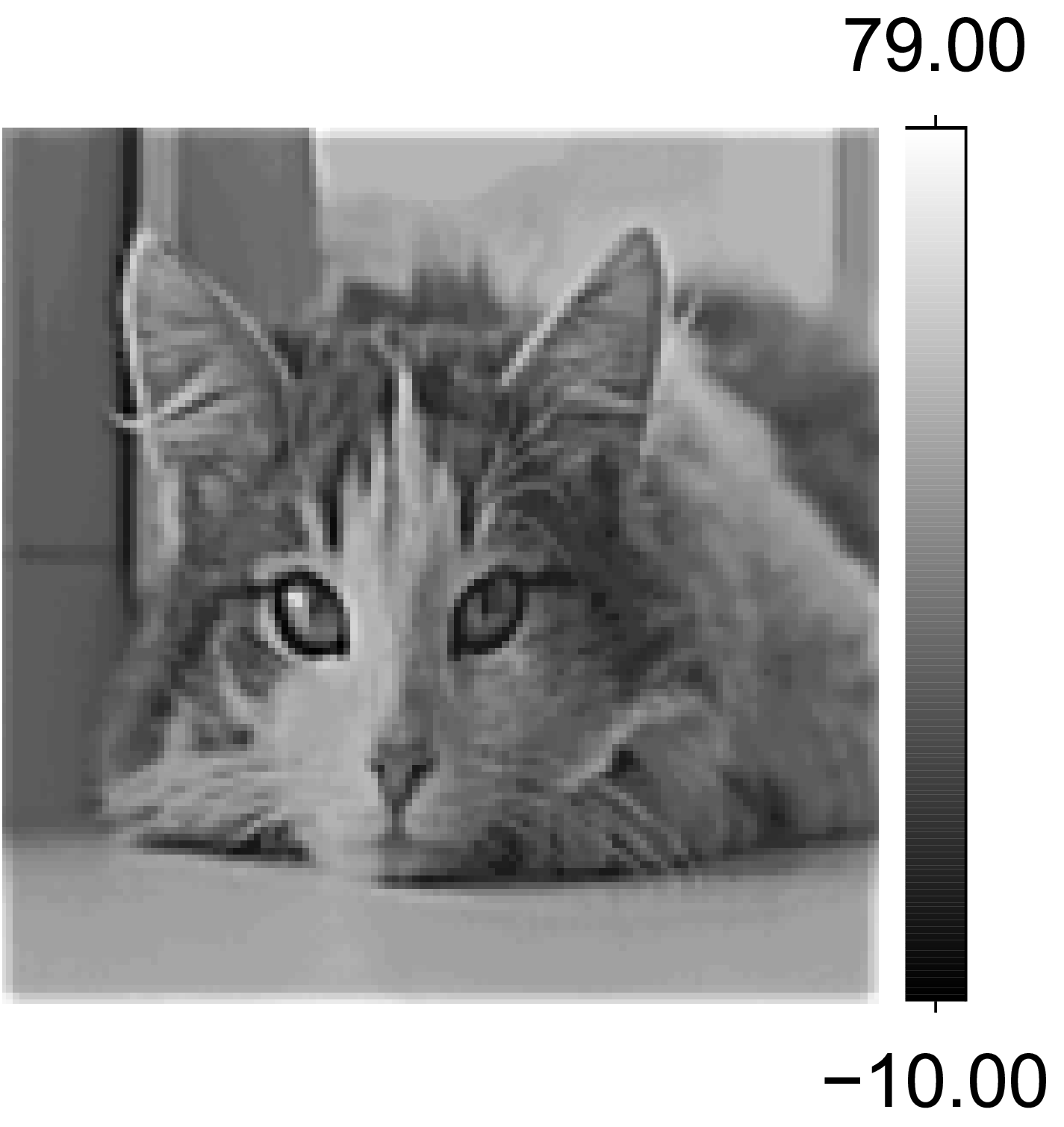}
		\label{bipolar}}		
  \subfloat[]{\includegraphics[width=0.247\linewidth]{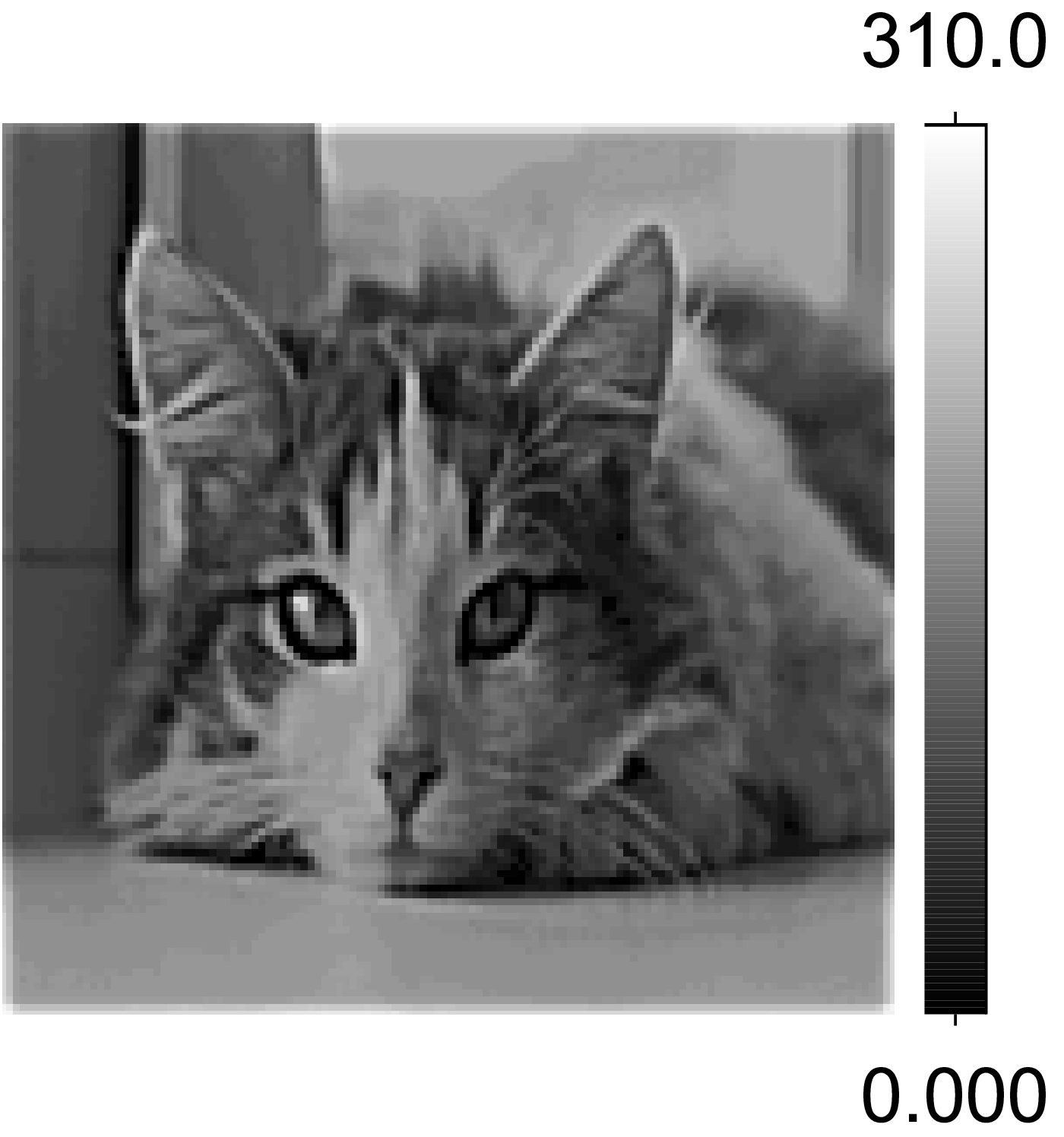}
		\label{giinput}}
	\subfloat[]{\includegraphics[width=0.247\linewidth]{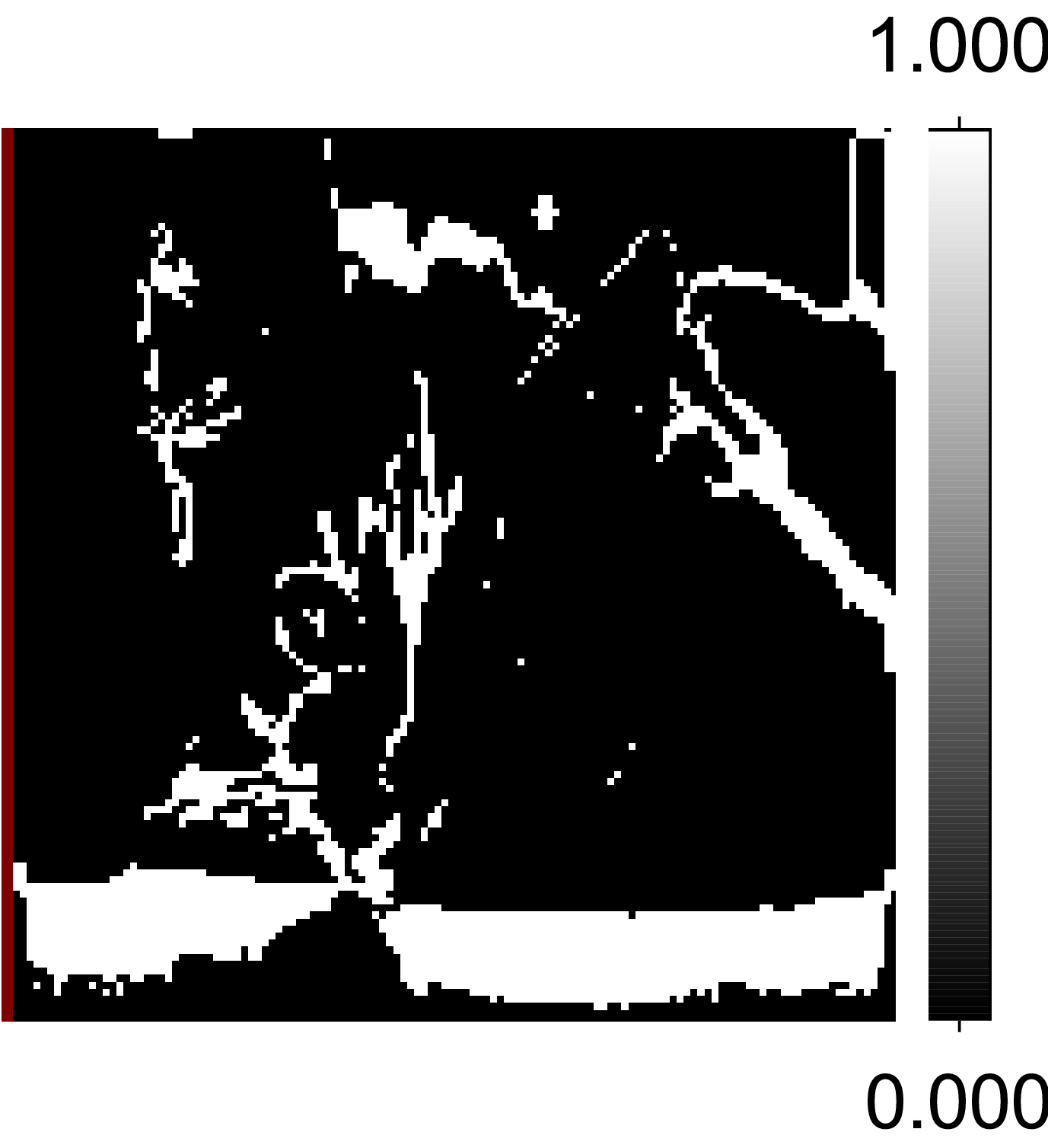}
		\label{spikeinputa}}
     \subfloat[]{\includegraphics[width=0.247\linewidth]{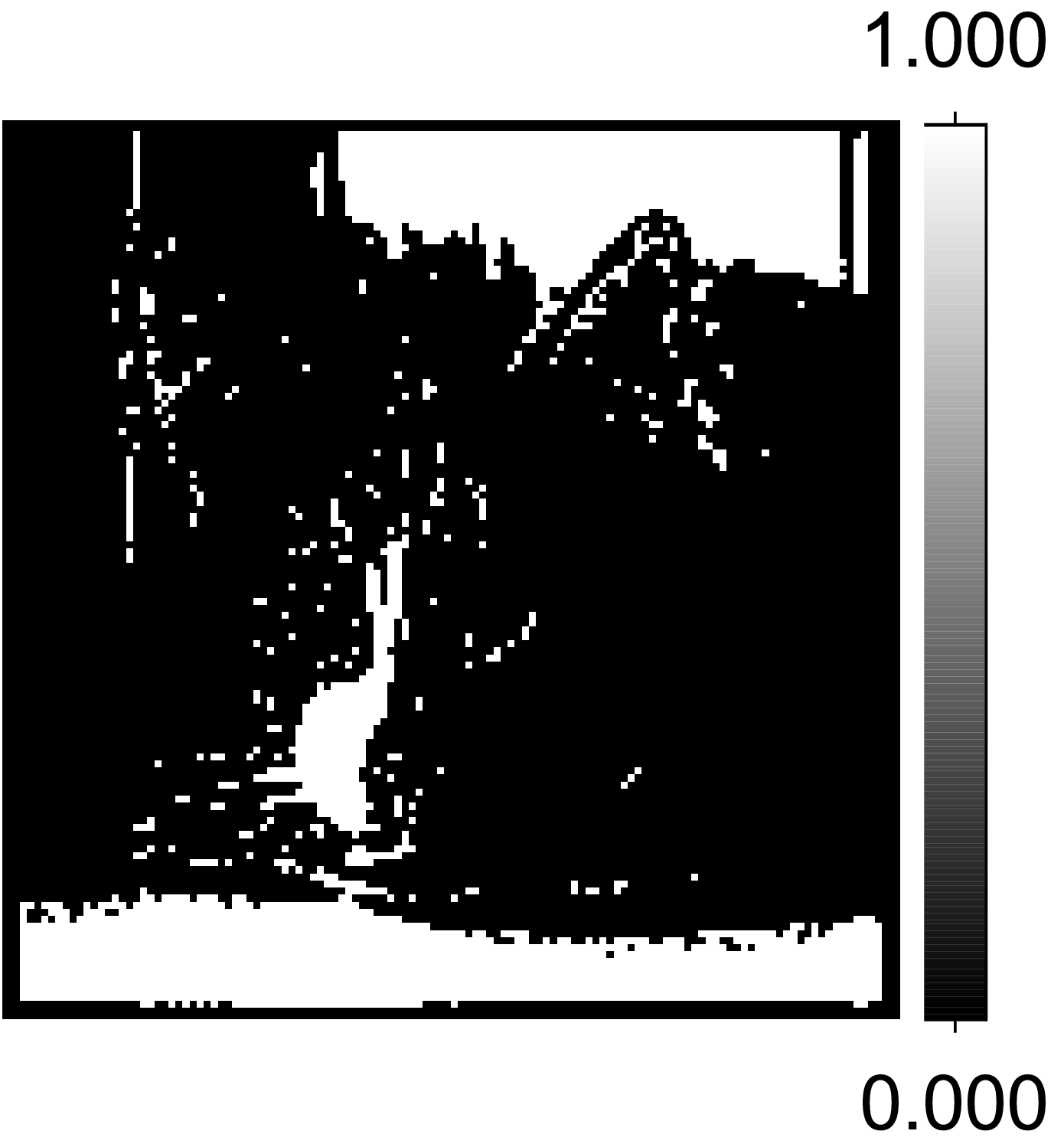}
		\label{spikeinputb}}\\
       \vspace{-0.10cm}
 \caption{The effect of the 10th frame of a video stimulus input on the different layers of the digital retina is illustrated; \protect\subref{imagein} Shows the input video stimulus frame;
\protect\subref{centerimage} Shows the center signal ($C$);
\protect\subref{surroundimage} Shows the surround signal ($S$);
\protect\subref{opljan} Shows the OPL layer output frame ($I_\text{OPL}$) from the digital retina;
\protect\subref{bipolar} Shows the bipolar layer output frame ($V_\text{Bip}$) from the digital retina;
\protect\subref{giinput} Shows the excitatory current on ganglion cells ($I_\text{Gang}$);
\protect\subref{spikeinputa} Depicts the ON spikes output frame from the digital retina;
\protect\subref{spikeinputb} Shows the OFF spikes output frame from the digital retina.}

	\label{imageout}
 
\end{figure*}
 
\begin{figure*}[!t]%
    \centering
    \subfloat[]{\includegraphics[width=0.245\linewidth]{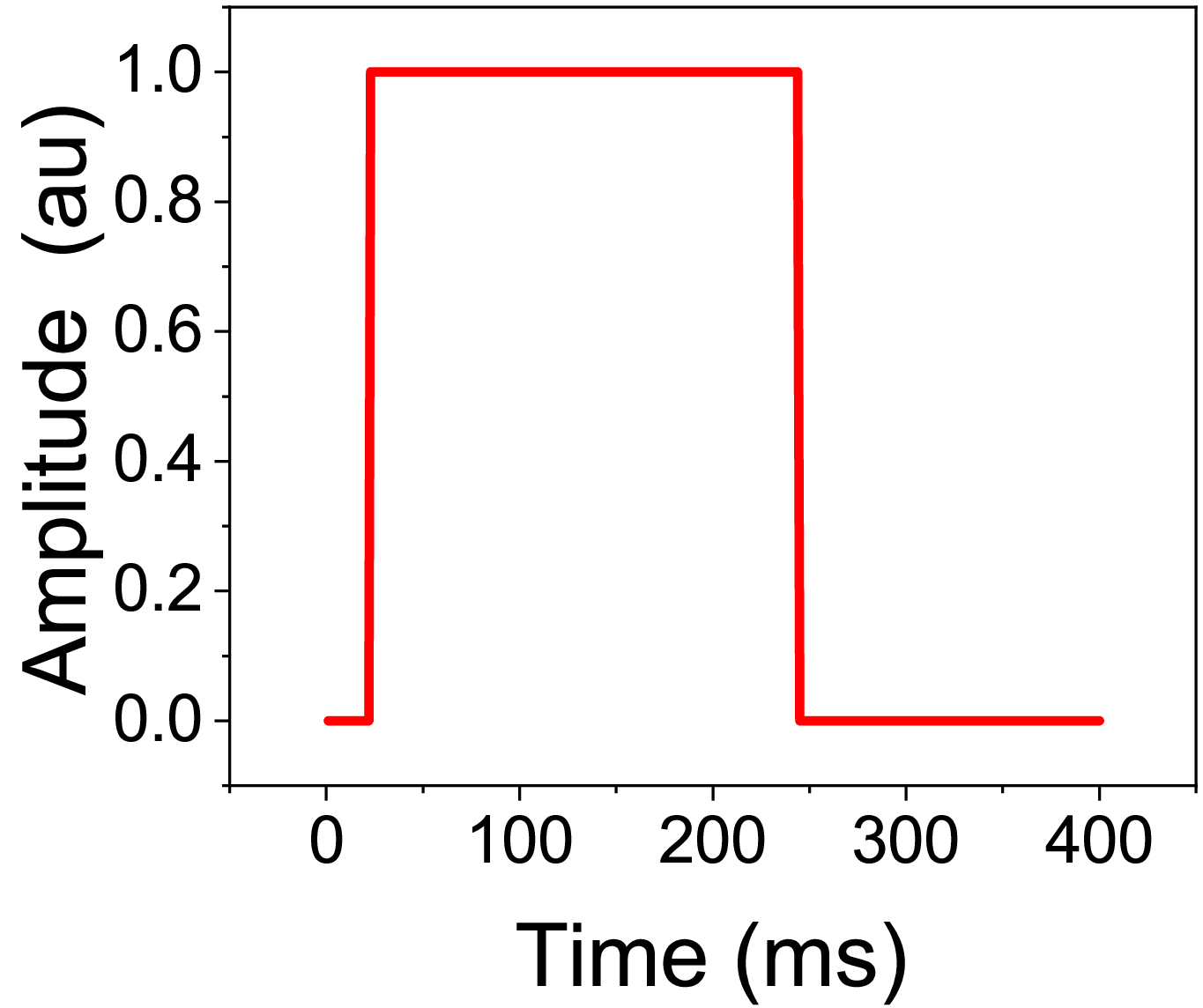}
    \label{pulseinput}}
     \subfloat[]{\includegraphics[width=0.25\linewidth]{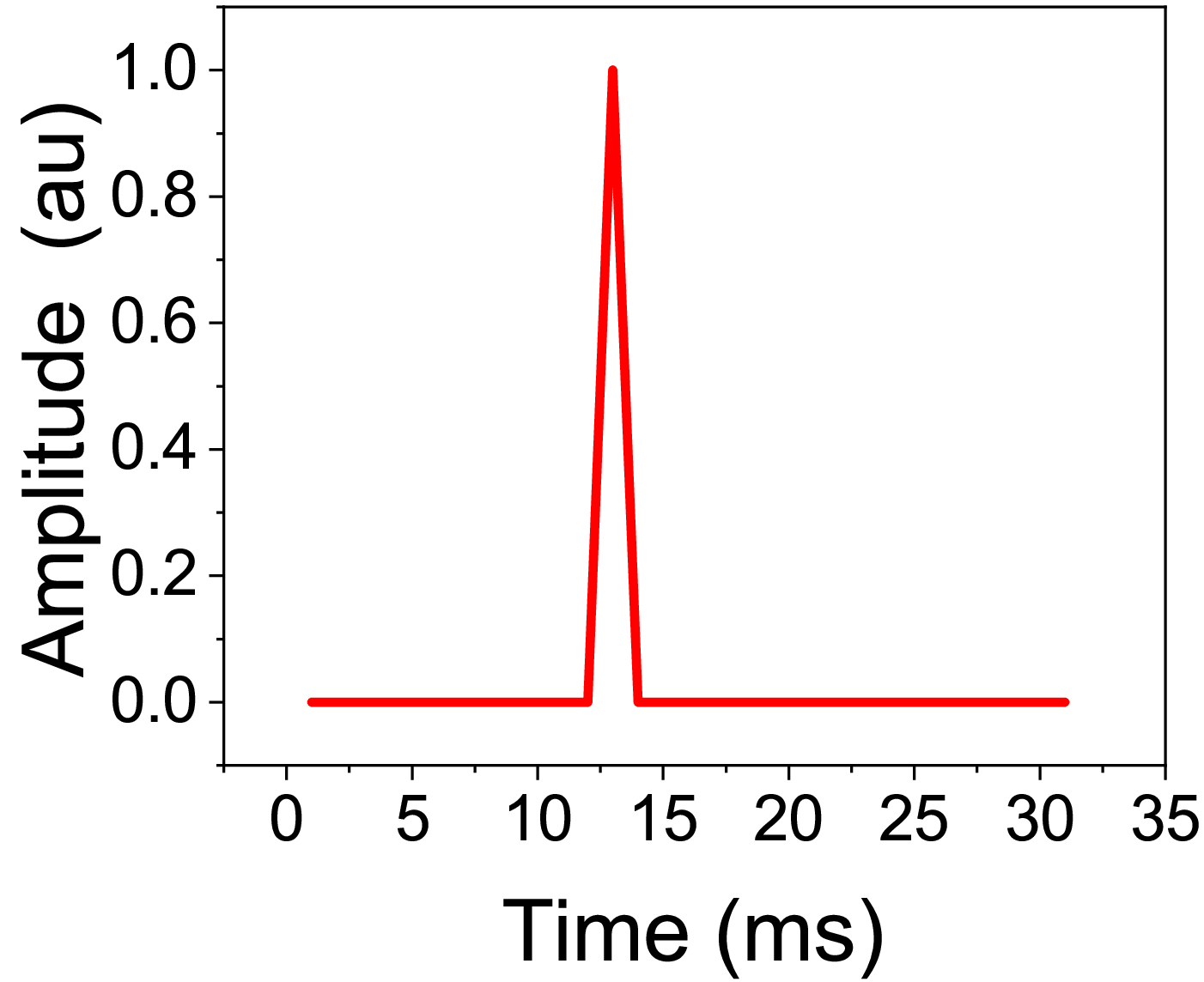}
    \label{singlespike}}
  \subfloat[]{\includegraphics[width=0.245\linewidth]{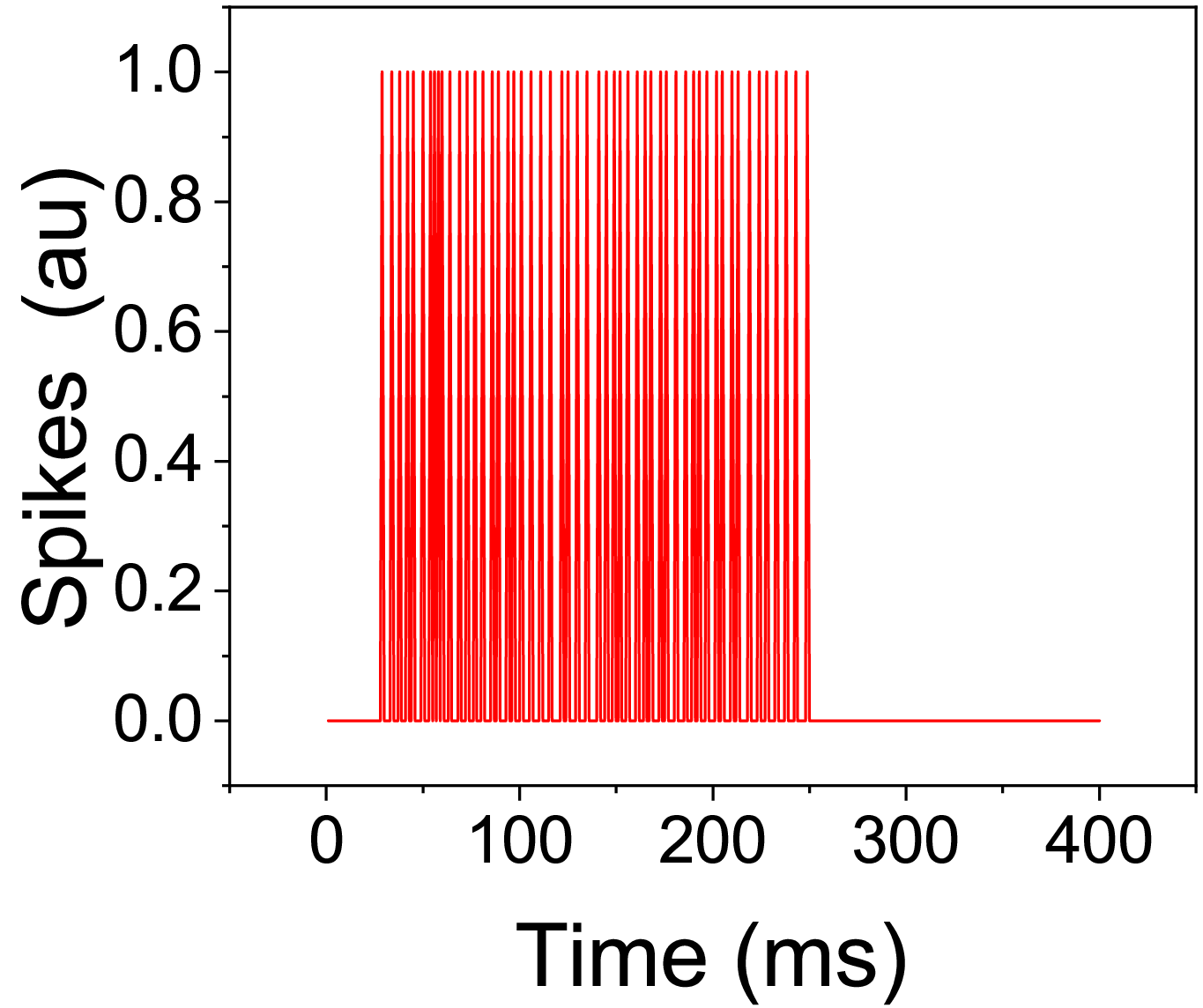}
    \label{tonic}}
    \subfloat[]{\includegraphics[width=0.245\linewidth]{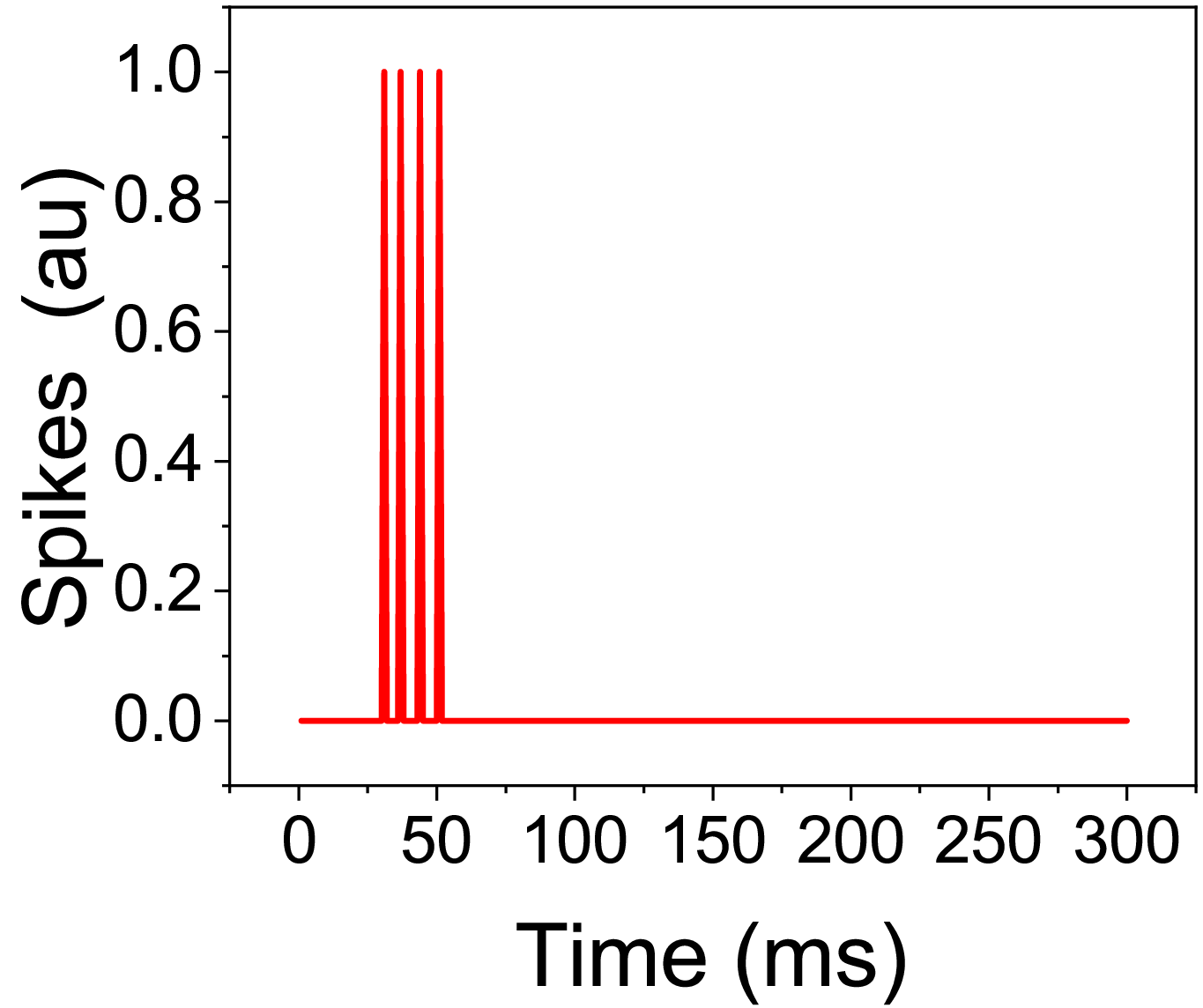}
    \label{phasic}}       
       \caption{Tonic and phasic responses of the retinal ganglion cell when subjected to a pulse signal; \protect\subref{pulseinput} Shows the pulse input applied at the photoreceptor cell of the retina; \protect\subref{pulseinput} Depicts the single spike output of the ganglion cell; \protect\subref{tonic} Shows the tonic response of the ganglion cell; \protect\subref{phasic} Shows the phasic cell response of the ganglion cell.}%
    \label{tonicphasic}%
\end{figure*}
The output of the center spatial filter ($G_\text{C}$) is then processed by the temporal low-pass filter ($E_{\tau \text{C}}$). To generate an output, the low-pass filter utilizes the previous output along with the present input.  A single $128\times128$ dual-port RAM is used to store the filter output, which is then used for processing the next frame. The low-pass filter is implemented using multiplier and adder structure and generates one output ($E_{\tau \text{C}}$) per clock cycle by taking input ($G_\text{C}$) and the previous frame output as input and storing the output in the same RAM.

The output from the low-pass filter is then passed through the temporal high-pass filter to obtain the center signal ($C$). The RAM associated with the high-pass filter is employed to store the previous frame information. The center signal is then processed by the surround spatial filter with a window size of 5$\times$5. The structure of the 5$\times$5 spatial filter closely resembles that of the 3$\times$3. Illustrated in Fig. \ref{conv55}, the register bank architecture for executing the 5$\times$5 convolution operation is depicted. This architecture employs four RAMs to store individual rows, with the numerical values inside the RAMs denoting the respective row indices. The allocation of rows to specific RAMs is indicated in Fig. \ref{conv55}.

The content of four RAMs along with 
the output of the surround spatial filter is then processed through another IIR low-pass filter to obtain the surround signal, which is subtracted from the center signal to obtain the OPL signal ($I_\text{OPL}$).
\subsection{ VLSI architecture of the bipolar layer} The architecture of the bipolar layer is shown in Fig. \ref{bipoarchi} and the detailed bipolar layer operations are defined by the pseudo-code in Table \ref{vb}. The OPL layer produces one output $I_\text{OPL}$ for each clock cycle, which is then used to generate the $g_\text{A}$ signal. This signal is further utilized to produce the $V_\text{Bip}$ signal by following the mathematical steps outlined in the pseudo-code. The hardware architecture for generating $V_\text{Bip}$ is simply the direct implementation of this pseudo-code. The $V_\text{Bip}$ signal is then used to generate the inhibition output, which is passed through the 5$\times$5 spatial filter window with the same architecture shown in Fig. \ref{conv55}. Ultimately, the $V_\text{Bip}$ signal serves as the final output from this layer.
\subsection{ VLSI architecture of the ganglion cells}
Table \ref{ig2} and \ref{spikes} present pseudo-codes that illustrate the spike generation process shown in Fig. \ref{bipoarchi}. Table \ref{ig2} outlines the generation process of the $I_\text{Gang}$ signal and provides a clear implementation for its VLSI architecture. This signal is subsequently used to generate the spike using the LIF neuron, which serves as the final output of the retina model. In generating the spike, the membrane potential $V_\text{m}$ is compared to a pre-defined threshold voltage, and a spike is generated when $V_\text{m}$ exceeds the threshold. The Artix-7 (Xa7a75tcsg324-2L), a 28nm technology node FPGA known for its low power consumption and high efficiency, was utilized to implement the proposed design. 
Input to the digital retina is a video input of size 128$\times$128 frames at 200fps. The maximum operating frequency of this design is 83MHz. This design consumes about 3720 LUTs, 3711 FFs, and 1720 Slices, along with 98.5 BRAMs (36 Kb), summarized in Table \ref{resource}. 
\section{Results and Discussion} 
\label{resultsdigital}

\begin{table*}[]
\resizebox{\textwidth}{!}{%
 \centering
  \begin{threeparttable}
 \caption{Comparison of this work with other digital retina models.}
\begin{tabular}{|c|c|c|c|c|c|}
\hline
\rowcolor[HTML]{C0C0C0} 
\textbf{Type} & \textbf{ \cite{nagy2005emulated}} & \textbf{ \cite{voroshazi2009fpga}} & \textbf{\cite{yang2019digital}} & \textbf{\cite{ghanbarpour2021efficient}} & \textbf{This Work} \\ \hline
\textbf{Year} & 2005 & 2009 & 2019 & 2021 & 2023 \\ \hline
\textbf{Model type} & CNN based & CNN based & modelled all layers & \begin{tabular}[c]{@{}c@{}}modelled only\\  photoreceptor cells\end{tabular} & \begin{tabular}[c]{@{}c@{}}mechanistic model\\ (modelled all \\ retinal layers)\end{tabular} \\ \hline
\textbf{Functionality} & Spatial & Spatial & Temporal & Temporal & Spatio-Temporal \\ \hline
\textbf{FPGA used} & Virtex-I1 6000 & \begin{tabular}[c]{@{}c@{}}Virtex-II \\ XC2V6000\end{tabular} & \begin{tabular}[c]{@{}c@{}}Cyclone IV \\ EP4CE115\end{tabular} & \begin{tabular}[c]{@{}c@{}}Virtex 5\\  XC5VLX20T\end{tabular} & \begin{tabular}[c]{@{}c@{}}Artix-7 \\ Xa7a75tcsg324-2L\end{tabular} \\ \hline
\textbf{Resource Utilization} & NA$^+$ & NA$^+$ & 5276(registers) & 8496 (Slices) & 1720 (Slices) \\ \hline
\textbf{Pixel array size} & 145$\times$145 & 256$\times$256 & NA$^+$ & NA$^+$ & 128$\times$128 \\ \hline
\textbf{Data path bit width} & 30 & 22-24 & NA$^+$ & 46 & 19 \\ \hline
\textbf{\begin{tabular}[c]{@{}c@{}}Maximum clock\\  frequency\end{tabular}} & 280MHz & 550MHz & NA$^+$ & 644.33MHz & 83MHz \\ \hline
\textbf{Frame rate} & 30fps & 25fps & NA$^+$ & NA$^+$ & 200fps$^*$ \\ \hline
\textbf{Contrast adaptation} & No & No & No & No & Yes \\ \hline
\textbf{Luminescence adaptation} & No & No & No & Yes & Yes \\ \hline
\textbf{\begin{tabular}[c]{@{}c@{}}Center-surround \\ receptive field\end{tabular}} & No & No & No & No & Yes \\ \hline
\textbf{\begin{tabular}[c]{@{}c@{}}Phasic cells and \\ tonic cells\end{tabular}} & No & No & No & No & Yes \\ \hline
\textbf{Ganglion cell type} & NA$^+$ & NA$^+$ & ON and OFF & No & ON and OFF \\ \hline
\end{tabular}
  \begin{tablenotes}
      \small
      \item + The relevant values for their designs were absent in these works.
      \item * The design can support 512$\times$512 pixel images at this frame rate.
    \end{tablenotes}
      \label{comparison}
  \end{threeparttable}
  }
\end{table*}

Fig. \ref{psf} illustrates how the photoreceptor cell responds to an impulse input, similar to the observation made by Schnapf \cite{schnapf1990visual} in the macaque retina. To ensure that the implementation of the digital retina matches the software model, a variety of stimuli and retinal combinations were used in one-to-one comparisons.
In order to guarantee that the temporal responses remain congruent with the identical model used in \cite{huth2017retina}, the properties of retinal cells are characterized using the chirp stimulus presented in Fig. \ref{chirpstimulus}. A quantized version of the Convis retina model has been implemented in MATLAB for software simulations and verifications. The FPGA design translates the MATLAB code utilizing fixed-point arithmetic.

The chirp stimulus shown in Fig. \ref{chirpstimulus} is used as the luminance input ($L$) for the digital retina, and the output signals of different layers are compared with the software model. Parameters are consistent between the simulations. Fig. \ref{oplchirp} illustrates the comparisons between software and hardware results of the OPL, bipolar, and ganglion excitatory current, as well as the spiking layers. Digital retina outputs are indicated in red, while software outputs are indicated in black. The high-pass filter effect can be observed in Fig. \ref{opla} and \ref{gia} with peaks at pulse edges. Rectification is evident in Fig. \ref{gia}. The digital retina can mimic the behavior of 'ON' and 'OFF' ganglion cells by adjusting the parameter $\xi$ to 1 or -1. Although biological processes differentiate 'ON' and 'OFF' cells earlier in retinal processing, the current model assumes a single and symmetrical signal until the bipolar cell stage without considering specific distinctions in reaction time, contrast gain control sensitivity, or cell density \cite{wohrer2008model}. Fig. \ref{oplchirp} illustrates spike plots for 'ON' and 'OFF' ganglion cells, demonstrating the closer resemblance of the digital realm to software retina responses. 

The variance explained by the software retina model at each stage of the digital retina is quantified in Fig. \ref{variance}. The OPL layer exhibits slight numerical discrepancies due to imprecise filters, but the bipolar and ganglion cell stages accurately replicate the software model findings. Cross-correlation of signals is used to determine the time difference between the digital retina and software, as shown in Fig. \ref{coor2}. The occurrence of a peak at the 0-time step signifies zero timing difference. In Fig. \ref{imageout}, the effect of a video stimulus at various retina layers is depicted, employing a frame size of 128$\times$128 pixels. Tonic and phasic responses are plotted in Fig. \ref{tonicphasic}. Tonic cells exhibit a continuous response as long as the stimulus is present (Fig. \ref{tonic}), while phasic cells show a brief response (Fig. \ref{phasic}).

A comparison of the design with state-of-the-art implementations is provided in Table \ref{comparison}. The design implements most of the retinal neurons and their properties as compared to other digital models. This table shows the retina includes both adaptation to luminance and contrast gain control as compared with other retinas.  These mechanisms are important as they allow the retina to accept a wide input range of the visual system. It can be seen from the table that the retina is reconfigurable, a crucial aspect for testing or comparing other biological retinas. 
The table shows that tonic and phasic cells are realizable compared to the state-of-the-art. Tonic cells provide a stable representation of a stimulus, while phasic cells are better suited for detecting changes in that stimulus.
 As a result, the visual system can recognize and react to various visual stimuli, including low-contrast and quickly changing objects, as well as high-contrast and slowly moving things. The digital retina processes 128$\times$128 pixel images at a frame rate of 200fps which is much superior as compared to other state-of-the-art models. Fewer hardware resources are utilized for this bio-plausible retina design. This demonstrates that the system can be implemented using minimal area, making it suitable for designing retinal prosthesis systems. Apart from retinal prostheses, the digital retina with these retinal characteristics may be of assistance to neuroscientists who need bio-plausible input for visual cortex therapies and robot vision. 

\section{Conclusion}
\label{conclusiondigital}
The main contribution of this work is the implementation of a spatio-temporal neuromorphic digital retina with key retinal properties that represent a concrete step towards building real-time biologically 
plausible retinal prostheses. The OPL filter of this retina serves both as an edge and movement detector. This digital retina incorporates the luminance adaptation of the photoreceptors. A shunt feedback mechanism is utilized to implement contrast gain control in the digital retina, which enables it to adapt to different levels of light contrast. Having both phasic and tonic cells in the retina imparts the advantage of offering a broad spectrum of visual processing capabilities. Additionally, temporal filtering and rectification in the IPL, as well as spike generation in ganglion cells, culminate in a biologically accurate output. This high-performance digital retina is capable of computing 512$\times$512 pixel frames at the frame rate of 200fps. This retina implementation on the FPGA provides a hardware platform for real-time visual data processing, which is crucial for applications in robotics, autonomous systems, and image processing. Furthermore, this project offers valuable insights into the human retina, potentially contributing to the development of new medical treatments for visual disorders.

\section*{Acknowledgment}
The authors would like to acknowledge the joint Memorandum of Understanding (MoU) between Indian Institute of Science, Bangalore, and the International Centre for Neuromorphic Systems, Western Sydney University, Australia. This work is supported by Pratiksha Trust.
\section*{Conflict of Interest Statement}
The authors have no conflicts of interest to disclose.

\FloatBarrier
\bibliographystyle{elsarticle-num}
\bibliography{ref}
\FloatBarrier

\vfill

\end{document}